\documentclass[aps,prd,preprintnumbers,groupedaddress,nofootinbib,amssymb,eqsecnum,notitlepage,dvipdfmx]{revtex4}
\usepackage{here}
\usepackage{graphicx}
\usepackage{amsmath,amsthm,amssymb}
\usepackage{bm}
\usepackage{color}

\usepackage{amsfonts}
\usepackage{dcolumn}
\usepackage{hyperref}
\allowdisplaybreaks[1]
\usepackage{ulem}

\begin{document}
\newcommand{\newc}{\newcommand}

\newcommand{\rk}{\textcolor{red}}
\newc{\be}{\begin{equation}}
\newc{\ee}{\end{equation}}
\newc{\ba}{\begin{eqnarray}}
\newc{\ea}{\end{eqnarray}}
\newc{\D}{\partial}
\newc{\rH}{{\rm H}}
\newc{\rd}{{\rm d}}

\title{Relativistic star perturbations 
in Horndeski theories with a gauge-ready formulation}

\author{Ryotaro Kase$^{1}$ and Shinji Tsujikawa$^{2}$}

\affiliation{
$^1$Department of Physics, Faculty of Science, 
Tokyo University of Science, 1-3, Kagurazaka,
Shinjuku-ku, Tokyo 162-8601, Japan\\
$^2$Department of Physics, Waseda University, 3-4-1 Okubo, Shinjuku, Tokyo 169-8555, Japan}

\begin{abstract}
We present a general framework for studying the relativistic star perturbations 
on a static and spherically symmetric background in full Horndeski theories.
We take a perfect fluid into account as a form of the Schutz-Sorkin action. 
Our formulation is sufficiently versatile in that the second-order actions of 
perturbations in odd- and even-parity sectors are derived without choosing particular gauge 
conditions, so they can be used for any convenient gauges at hand.  
The odd-parity sector contains one dynamical gravitational degree 
of freedom coupled to a time-independent four velocity of the fluid. 
In the even-parity sector there are three dynamical perturbations associated 
with gravity, scalar field, and matter sectors, whose equations of motion
are decoupled from other nondynamical perturbations. 
For high radial and angular momentum modes, we obtain the propagation 
speeds of all dynamical perturbations and show that the perfect 
fluid in the even-parity sector has a standard sound speed affected 
by neither gravity nor the scalar field. Our general stability conditions and 
perturbation equations of motion can be directly applied to  
the stabilities of neutron stars and black holes 
as well as the calculations of their quasi-normal frequencies.

\end{abstract}

\date{\today}

%\pacs{04.50.Kd, 95.36.+x, 98.80.-k}

\preprint{WUCG-21-13}

\maketitle

%%%%%%%%%%%%%%%%%%%%%%%%%%%%%%%%%%%%%%%%%%
\section{Introduction}
\label{introsec}
%%%%%%%%%%%%%%%%%%%%%%%%%%%%%%%%%%%%%%%%%%

The physics of black holes (BHs) and neutron stars (NSs) can be probed 
by observing their binary systems through 
gravitational waves (GWs) \cite{Abbott2016,GW170817}.
Precise measurements of the GW waveform of a compact binary 
coalescence event yield a plethora of information about the accuracy 
of General Relativity (GR) and their possible 
deviations on a strong gravitational background 
\cite{Cutler:1994ys,Vitor1,Pani,Yagi1,Yagi2,Blanchet:2013haa,Berti:2015itd,Barack}.  
Moreover the tidal deformation of a NS-NS binary system provides useful 
constraints on their mass-radius 
relation \cite{Flanagan:2007ix,Hinderer:2007mb,Hinderer:2009ca}, 
so that the NS equations of state are probed by 
GW measurements \cite{LIGOScientific:2018cki,De:2018uhw}. 
After the merging of binaries the BHs and NSs exhibit quasi-normal 
oscillations \cite{Kokkotas:1999bd,Nollert,Berti:2009kk}, 
whose frequencies can be computed by 
studying perturbations in the odd- and even-parity sectors. 

GR is a fundamental theory of gravity whose accuracy has been confirmed 
on the weak gravitational background \cite{Will}. 
However, it is yet unknown how much extent GR accurately describes 
the physics around strong gravitational objects.
On the cosmological side, there are several unsolved 
problems like the origin of dark energy, dark matter, inflation, 
and the matter-antimatter asymmetry. 
The existence of such problems may imply the presence of 
extra degrees of freedom (DOFs) beyond those appearing in 
GR and standard model of particle 
physics \cite{CST,Silvestri,Clifton,Joyce,Kase:2018aps}. 
Those extra DOFs can be a scalar field, vector field, or a massive 
graviton, and so on. Among them, the scalar field can be naturally 
accommodated on an isotropic and homogenous cosmological background, 
so it has been extensively applied 
to the physics of inflation and dark energy. 

The theories in which a scalar field $\phi$ has nonminimal or
derivative couplings to gravity are generally called 
scalar-tensor theories \cite{Fujii}. 
In particular, Horndeski theories \cite{Horndeski} are the 
most general scalar-tensor theories with second-order field 
equations of motion (see also Refs.~\cite{Def11,KYY,Charmousis:2011bf}). 
In subclasses of Horndeski theories, there have been many works 
for the search of ``hairy'' BH and NS solutions 
on a static and spherically symmetric background.
In GR with a minimally coupled scalar field, there is a no-hair BH 
theorem stating that the BHs are described by only three parameters, 
i.e., mass, electric charge, and angular 
momentum \cite{Chase,BekenPRL}. 
This property also holds for a nonminimally coupled scalar field 
with the Ricci scalar $R$ \cite{Hawking72,Beken95,Soti12} 
(e.g., Brans-Dicke theories \cite{Brans:1961sx}). 
In the presence of a Gauss-Bonnet term linearly coupled to $\phi$, 
there exists a hairy BH solution where the 
background geometry is affected by the scalar field \cite{Soti1,Soti2}. 
If we do not impose the asymptotic flatness, nonminimal derivative 
couplings to the Einstein tensor $G_{\mu \nu}$ give rise to hairy BH solutions 
with a nonvanishing scalar-field profile 
$\phi(r)$ \cite{Rinaldi:2012vy,Minamitsuji:2013ura,Anabalon:2013oea}.
It is also known that a time-dependent scalar field 
of the form $\phi=qt+\psi(r)$ allows the existence of 
a nontrivial stealth Schwarzschild 
solution \cite{Babi14,Koba14,Ogawa:2015pea,Babichev:2016rlq,Khoury:2020aya}.

In scalar-tensor theories with a nonminimal coupling of the form 
$F(\phi)R$, there exist some static and spherically symmetric 
NS solutions with scalar 
hairs \cite{Damour,Damour2,Cooney:2009rr,Arapoglu:2010rz,Orellana:2013gn,
Astashenok:2013vza,Ganguly,Yazadjiev:2014cza,Resco:2016upv,Kase:2019dqc}. 
For the coupling $F(\phi)$ containing even power-law functions of $\phi$, 
there is a nontrivial hairy branch ($\phi \neq 0$) besides the standard 
general relativistic branch ($\phi=0$ everywhere).
If the second derivative $\rd^2 F/\rd \phi^2$ is positive at $\phi=0$, 
the GR branch can be unstable to trigger tachyonic 
growth of $\phi$ toward the hairy branch.
This phenomenon, which is induced by a tachyonic mass of the scalar field at $\phi=0$, 
is dubbed spontaneous scalarization \cite{Damour,Damour2}.
In the presence of a Gauss-Bonnet term coupled to the scalar field,  
it has been recognized that spontaneous scalarization 
can also occur for nonrotating and rotating 
BHs \cite{Doneva:2017bvd,Silva:2017uqg,Antoniou:2017acq,
Antoniou:2017hxj,Minamitsuji:2018xde,Cunha}. 

Nonminimally coupled theories with the coupling $F(\phi)R$ 
include Brans-Dicke theories with a scalar potential and $f(R)$ gravity.
In these theories the nonminimal coupling $F(\phi)$ 
is not given by even power-law functions of $\phi$, 
so spontaneous scalarization does not occur. 
Applying the $f(R)$ models of late-time cosmic acceleration to relativistic 
stars \cite{Frolov:2008uf,Kobayashi:2008tq,Tsujikawa:2009yf,Upadhye:2009kt}, 
the existence of a growing scalar mass in regions of high density 
can give rise to a rapidly increasing solution of the scalar degree of freedom. 
In this case, the fine tuning of boundary conditions 
is required at the center of spherical symmetry. 
NS solutions in Starobinsky model \cite{Starobinsky}, which is 
described by the Lagrangian $f(R)=R+R^2/(6m^2)$ with a constant mass $m$, 
were studied in Ref.~\cite{Yazadjiev:2014cza} by using a non-perturbative 
approach both inside and outside the NS.
(see Refs.~\cite{Cooney:2009rr,Arapoglu:2010rz} for a perturbative 
approach). There are also models with the Lagrangian 
$f(R)=R+aR^p$, where $p$ is a constant in the range 
$1<p<2$. In this case, the effective scalar mass 
depends on the field value and approaches 0 toward spatial infinity. 
NS solutions of this class of models and its cooling process were studied 
in Refs.~\cite{Kase:2019dqc,Dohi:2020bfs}. 
In a subclass of derivative coupling theories where the scalar field is coupled to 
an Einstein tensor $G_{\mu \nu}$ of the form 
$\phi G_{\mu \nu}\nabla^{\mu}\nabla^{\nu}\phi$, there are also 
nontrivial NS solutions endowed with a scalar hair only 
inside the star \cite{Cisterna:2015yla,Cisterna:2015uya,Maselli}.

To study the stabilities of hairy BHs and NSs in scalar-tensor theories, 
we need to consider the perturbations on a static and spherically symmetric 
background. For BHs, this issue was addressed in full Horndeski theories for odd-parity 
modes \cite{Kobayashi:2012kh} and even-parity 
modes \cite{Kobayashi:2014wsa} (see 
Refs.~\cite{Tattersall:2018nve,Franciolini:2018uyq,Glampedakis:2019dqh,
Tattersall:2019nmh,Tomikawa:2021pca,Langlois:2021aji} 
for the application to quasinormal modes). 
For NSs, the analysis of perturbations with matter was carried out 
in Ref.~\cite{Kase:2020qvz} in a subclass of Horndeski theories given by 
the Lagrangian ${\cal L}=G_2(\phi, X)+G_4(\phi)R$. 
The stability of derivative coupling theories containing the Lagrangian 
$\phi G_{\mu \nu}\nabla^{\mu}\nabla^{\nu}\phi$ was also performed 
in Ref.~\cite{Kase:2020yjf}. 
However, the general stability conditions of NSs in full Horndeski 
theories have not been derived yet.

The Lagrangian of full Horndeski theories contains the cubic self-interaction
$G_3(\phi, X) \square \phi$ as well as nonminimal and derivatives 
couplings to gravity, see Eq.~(\ref{LH}) in Sec.~\ref{scasec}.
On the expanding cosmological background, the Lagrangian of theories 
with the speed of gravitational waves $c_t$ exactly equivalent to that of 
light ($c$) is constrained to be of the form 
${\cal L}=G_2(\phi, X)+G_3(\phi, X) \square \phi
+G_4(\phi)R$. 
From the gravitational wave event GW170817 \cite{GW170817} 
together with the $\gamma$-ray observation of its electromagnetic 
counterpart \cite{Goldstein}, there is a constraint 
$-3\times10^{-15}\leq c_t-1\leq7\times10^{-16}$ 
in the unit of $c=1$. The nonminimal derivative coupling 
$G_5(\phi) \propto \phi$ and linear Gauss-Bonnet coupling 
$G_5(X) \propto \ln X$ lead to the deviation of $c_t$ from 1. 
However, the deviation usually occurs only in the vicinity of strong 
gravitational objects, so the BH and NS solutions in Horndeski theories 
containing the couplings $G_4(X)$ and $G_5(\phi, X)$ are not 
necessarily excluded from the observational bound of speed of 
gravitational waves.

In this paper, we will compute the second-order actions of odd- and even-parity
perturbations in full Horndeski theories in the presence of a perfect fluid 
described by the Schutz-Sorkin action. 
We do not fix gauge conditions from the beginning, so the resulting 
perturbation equations of motion can be applied to any gauges. 
We will derive stability conditions of relativistic stars against 
odd- and even-parity perturbations for high radial and angular 
momentum modes.  For BHs, i.e., in the absence of the matter fluid, 
the similar investigations were performed in 
Refs.~\cite{Kobayashi:2012kh,Kobayashi:2014wsa}, but 
the stability of even-parity perturbations along the angular direction 
was not addressed. In the presence of matter, we will obtain 
all the stability conditions including those in the angular direction.
Thus, our analysis is sufficiently general to accommodate the 
stabilities of NSs as well as BHs.

For the integration of the matter action, the prescriptions taken in 
Refs.~\cite{Kase:2020qvz,Kase:2020yjf} contain some inappropriate 
points. In this paper, we will also correct them and show that the dynamical matter perturbation 
in the even-parity sector propagates with the standard sound speed $c_m$
affected by neither gravity nor the scalar field. 
We will also apply our results to derivative coupling theories 
containing the Lagrangian $\phi G_{\mu \nu}\nabla^{\mu}\nabla^{\nu}\phi$ 
and show that the main results of instabilities of hairy solutions in this class of 
theories \cite{Kase:2020yjf} are not subject to modifications. 
We would like to stress that our general gauge-ready formulation of 
relativistic star perturbations in 
full Horndeski theories can be directly applied to the computations of 
quasi-normal modes of NSs and BHs as well as to the tidal deformation 
of NSs.

%%%%%%%%%%%%%%%%%%%%%%%%%%%%%%%%%%%%%%%%%%
\section{Horndeski theories with matter and background equations 
on a static and spherically symmetric background}
\label{scasec}
%%%%%%%%%%%%%%%%%%%%%%%%%%%%%%%%%%%%%%%%%%

The action of most general scalar-tensor theories with second-order field 
equations of motion is given by \cite{Horndeski,Def11,KYY}
\be
{\cal S}=\int {\rm d}^4 x \sqrt{-g}\,{\cal L}_H
+{\cal S}_m (g_{\mu \nu}, \Psi_m)\,,
\label{action}
\ee
where $g$ is a determinant of the metric tensor $g_{\mu \nu}$, and 
\ba
{\cal L}_H
&=&
G_2(\phi,X)-G_{3}(\phi,X)\square\phi 
+G_{4}(\phi,X)\, R +G_{4,X}(\phi,X)\left[ (\square \phi)^{2}
-(\nabla_{\mu}\nabla_{\nu} \phi)
(\nabla^{\mu}\nabla^{\nu} \phi) \right]
+G_{5}(\phi,X)G_{\mu \nu} \nabla^{\mu}\nabla^{\nu} \phi
\notag\\
&&
-\frac{1}{6}G_{5,X}(\phi,X)
\left[ (\square \phi )^{3}-3(\square \phi)\,
(\nabla_{\mu}\nabla_{\nu} \phi)
(\nabla^{\mu}\nabla^{\nu} \phi)
+2(\nabla^{\mu}\nabla_{\alpha} \phi)
(\nabla^{\alpha}\nabla_{\beta} \phi)
(\nabla^{\beta}\nabla_{\mu} \phi) \right]\,,
\label{LH}
\ea
with $R$ and $G_{\mu \nu}$ being the Ricci scalar and 
Einstein tensor, respectively.
The four functions $G_{j}$'s ($j=2,3,4,5$) depend on  
the scalar field $\phi$ and its kinetic term 
$X=-g^{\mu\nu}\nabla_{\mu}\phi\nabla_{\nu}\phi/2$, 
with the covariant derivative operator $\nabla_{\mu}$. 
We also use the notations  
$\square \phi \equiv \nabla^{\mu}\nabla_{\mu} \phi$ and 
$G_{j,\phi} \equiv \partial G_j/\partial \phi$, 
$G_{j,X} \equiv \partial G_j/\partial X$, 
$G_{j,\phi X} \equiv \partial^2 G_j/(\partial X \partial \phi)$,  etc. 

For the matter fields $\Psi_m$ inside relativistic stars, 
we consider a perfect fluid minimally coupled to gravity.
This is described by the Schutz-Sorkin 
action \cite{Sorkin,Brown,DGS,DeFelice:2016yws} 
\be
{\cal S}_{m} =  -\int {\rm d}^{4}x \left[
\sqrt{-g}\,\rho(n)
+ J^{\mu} (\partial_{\mu} \ell+{\cal A}_i\partial_{\mu}{\cal B}^i)\right]\,,
\label{SM}
\ee
where $\rho$ is a matter density depending on its number density $n$, 
and $J^{\mu}$ is a vector field related to the fluid four 
velocity $u^{\mu}$ as 
\be
u^{\mu}=\frac{J^{\mu}}{n\sqrt{-g}}\,.
\label{defu}
\ee
Due to the relation $u^{\mu} u_{\mu}=-1$, the fluid number density 
$n$ is expressed as 
\be
n=\sqrt{\frac{g_{\mu \nu}J^{\mu} J^{\nu}}{g}}\,.
\label{defn}
\ee
The scalar field $\ell$ and spatial vectors 
${\cal A}_i$, ${\cal  B}^i$ ($i=1,2,3$) are the Lagrange multipliers 
arising in accompany with the current vector field $J^{\mu}$. 
In the matter action (\ref{SM}), we do not take 
an entropy mode into account.

\subsection{Variation of the matter action}

The covariant field equations of motion following from the 
Horndeski Lagrangian (\ref{LH}) are explicitly given in Ref.~\cite{KYY}, 
so we do not repeat to write them here. 
In the following, we briefly revisit the field equations in 
the perfect-fluid sector. 
Varying the Schutz-Sorkin action (\ref{SM}) with respect to 
$J^{\mu}$, $\ell$, ${\cal A}_i$, and ${\cal B}^i$, respectively, we obtain
\ba
\partial_{\mu} \ell &=&
\rho_{,n} u_{\mu}
-{\cal A}_i\partial_{\mu}{\cal B}^i\,,
\label{rhomu}\\
\partial_{\mu} J^{\mu} &=& 0\,,
\label{Jcon}\\
J^{\mu} \partial_{\mu} {\cal B}^i &=&0\,,
\label{JAB1}\\
J^{\mu} \partial_{\mu} {\cal A}_i &=& 0\,,
\label{JAB2}
\ea
where $\rho_{,n} \equiv \partial \rho/\partial n$. 
For the derivation of Eq.~(\ref{JAB2}), we used the relation (\ref{Jcon}). 
Substituting $J^{\mu}=n \sqrt{-g}\,u^{\mu}$ into Eq.~(\ref{Jcon}), 
it follows that 
\be
u^{\mu}\nabla_{\mu}\rho
+(\rho+P)\nabla_{\mu}u^{\mu}=0\,, 
\label{umu}
\ee
where $P$ is the matter pressure defined by 
\be
P \equiv n \rho_{,n}-\rho\,.
\ee
The Lagrange multipliers $\ell$, ${\cal A}_i$, and ${\cal B}^i$ are
related to the physical quantities $\rho$ and $u^{\mu}$ 
through Eqs.~(\ref{rhomu}), (\ref{JAB1}), and (\ref{JAB2}). 
We will see that these Lagrange multipliers finally disappear from 
the background and perturbation equations 
of motion.

As shown in Refs.~\cite{Amendola:2020ldb,Kase:2020qvz}, 
varying the matter Lagrangian 
$L_{m}=-[\sqrt{-g}\,\rho(n)
+ J^{\mu} (\partial_{\mu} \ell
+{\cal A}_i\partial_{\mu}{\cal B}^i)]$
with respect to $g^{\mu \nu}$ 
leads to a familiar form 
of the perfect-fluid energy momentum tensor
\be
T_{\mu \nu} \equiv 
-\frac{2}{\sqrt{-g}} \frac{\delta L_m}{\delta g^{\mu \nu}}
=\left( \rho+P \right) u_{\mu} u_{\nu}+P g_{\mu \nu}\,,
\label{Tmunu}
\ee
where we used the relation (\ref{rhomu}).
Since the matter sector is minimally coupled to gravity, 
it obeys the continuity equation
\be
\nabla^{\mu} T_{\mu \nu}=0\,.
\label{Tcon}
\ee
Multiplying $u^{\nu}$ with Eq.~(\ref{Tcon}), the resulting equation 
is identical to Eq.~(\ref{umu}). 
Introducing a unit vector $n^{\nu}$ orthogonal to 
$u_{\nu}$ (i.e., $n^{\nu}u_{\nu}=0$ and $n^{\nu}n_{\nu}=1$) 
and multiplying $n^{\nu}$ with Eq.~(\ref{Tcon}), we obtain
\be
n_{\mu} \nabla^{\mu} P
=-(\rho+P)n^{\nu}u_{\mu} \nabla^{\mu} u_{\nu}\,.
\label{Tmueq2}
\ee
If we apply this relation to the configuration of compact stars, 
it can be interpreted as a balance between the pressure 
gradient and gravity \cite{Kase:2020qvz,Minamitsuji:2021vdb}.

\subsection{Background equations}
\label{backsec}

We consider a static and spherically symmetric  background 
given by the line element
\be
\rd s^2=-f(r) \rd t^{2} +h^{-1}(r) \rd r^{2}
+ r^{2} \left(\rd \theta^{2}+\sin^{2}\theta\,\rd\varphi^{2} 
\right)\,,
\label{BGmetric}
\ee
where $f(r)$ and $h(r)$ are functions of the distance $r$ from 
the center of symmetry. On this background, 
we consider the radial dependent scalar field,  
\be
\phi=\phi(r)\,.
\ee
For the matter sector, we can take the following 
configurations,
\be
J^{\mu}=\left[ \sqrt{-g}\,N(r),0,0,0 \right]\,, \qquad
{\cal A}_i=0\,,
\label{JBG}
\ee
where $\sqrt{-g}=f^{1/2}h^{-1/2}r^2\sin{\theta}$. 
{}From Eq.~(\ref{defn}), the quantity $N(r)$ is related to 
the fluid number density $n$, as 
\be
n(r)=f(r)^{1/2}N(r)\,.
\label{nBG}
\ee
Substituting Eqs.~(\ref{JBG}) and (\ref{nBG}) into Eq.~(\ref{defu}), 
the fluid four velocity is expressed in the form 
\be
u^{\mu}=\left[ f(r)^{-1/2},0,0,0 \right]\,.
\ee

Varying the action (\ref{action}) with respect to $g_{\mu\nu}$, 
the (00), (11), (22) components of gravitational field equations 
of motion are given, respectively, by 
\ba
{\cal E}_{00}&\equiv&
\left(A_1+\frac{A_2}{r}+\frac{A_3}{r^2}\right)\phi''
+\left(\frac{\phi'}{2h}A_1+\frac{A_4}{r}+\frac{A_5}{r^2}\right)h'
+A_6+\frac{A_7}{r}+\frac{A_8}{r^2}=\rho
\,,\label{back1}\\
{\cal E}_{11}&\equiv&
-\left(\frac{\phi'}{2h}A_1+\frac{A_4}{r}+\frac{A_5}{r^2}\right) \frac{hf'}{f}
+A_9-\frac{2\phi'}{r}A_1-\frac{1}{r^2}\left[\frac{\phi'}{2h}A_2+(h-1)A_4\right]=P
\,,\label{back2}\\
{\cal E}_{22}&\equiv&
\left[\left\{A_2+\frac{(2h-1)\phi'A_3+2hA_5}{h\phi' r}\right\}\frac{f'}{4f}+A_1+\frac{A_2}{2r}\right]\phi''
+\frac{1}{4f}\left(2hA_4-\phi'A_2+\frac{2hA_5-\phi'A_3}{r}\right)\left(f''-\frac{f'^2}{2f}\right)\notag\\
&&
+\left[A_4+\frac{2h(2h+1)A_5-\phi'A_3}{2h^2r}\right]\frac{f'h'}{4f}
+\left(\frac{A_7}{4}+\frac{A_{10}}{r}\right)\frac{f'}{f}
+\left(\frac{\phi'}{h}A_1+\frac{A_4}{r}\right)\frac{h'}{2}+A_6+\frac{A_7}{2r}
=-P\,,\label{back3}
\ea
where a prime represents the derivative with respect to $r$. 
The coefficients are given by
\ba
&&
A_1=-h^2 (G_{3,X}-2 G_{4,\phi X} ) \phi'^2-2 G_{4,\phi} h\,,\notag\\
&&
A_2=2 h^3 ( 2 G_{4,XX}-G_{5,\phi X} ) \phi'^3-4 h^2 ( G_{4,X}-G_{5,\phi} ) \phi'\,,\notag\\
&&
A_3=-h^4G_{5,XX} \phi'^4+h^2G_{5,X}  ( 3 h-1 ) \phi'^2\,,\notag\\
&&
A_4=h^2 ( 2 G_{4,XX}-G_{5,\phi X} ) \phi'^4+h ( 3 G_{5,\phi}-4 G_{4,X} ) \phi'^2-2 G_4\,,\notag\\
&&
A_5=-\frac12 \left[G_{5,XX} h^3{\phi'}^{5}- hG_{5,X}  ( 5 h-1 ) \phi'^3\right]\,,\notag\\
&&
A_6=h ( G_{3,\phi}-2 G_{4,\phi\phi} ) \phi'^2+G_2\,,\notag\\
&&
A_7=-2 h^2 ( 2 G_{4,\phi X}-G_{5,\phi\phi} ) \phi'^3-4 G_{4,\phi} h\phi'\,,\notag\\
&&
A_8=G_{5,\phi X} h^3\phi'^4-h ( 2 G_{4,X} h-G_{5,\phi} h-G_{5,\phi} ) \phi'^2-2 G_4  ( h-1 )\,,\notag\\
&&
A_9=-h ( G_{2,X}-G_{3,\phi} ) \phi'^2-G_2\,,\notag\\
&&
A_{10}=\frac12 G_{5,\phi X} h^3\phi'^4-\frac12 h^2 ( 2 G_{4,X}-G_{5,\phi} ) \phi'^2-G_4 h\,.
\ea
The matter continuity Eq.~(\ref{Tmueq2}) translates to
\be
{\cal E}_{P}\equiv P'+\frac{f'}{2f} \left( \rho+P \right)=0\,.
\label{back4}
\ee
Variation of the action (\ref{action}) with respect to $\phi$ 
leads to the scalar-field equation of motion 
\be
\frac{1}{r^2} \sqrt{\frac{h}{f}} \left( r^2 \sqrt{\frac{f}{h}} J^r 
\right)'+\frac{\partial {\cal E}}{\partial\phi}=0\,,
\label{Ephi}
\ee
where
\ba
J^r&=&\left(A_1+\frac{A_2}{r}+\frac{A_3}{r^2}\right)\frac{f'}{2f}
-\frac{A_6+A_9}{\phi'}+\frac{2}{r}A_1+\frac{1}{r^2}\left(\frac{1+h}{2h}A_2
-\frac{A_4+A_8-2A_{10}}{\phi'}\right)\,,\\
{\cal E}&=&\left[A_1+\frac{1}{r^2}\left(\frac{A_3}{2h}-\frac{A_5}{\phi'}\right)\right]
\left(\phi''+\frac{\phi'h'}{2h}\right)
+\left[\frac{\phi'}{2}A_2-hA_4+\frac{1}{2r}\left(\frac{\phi'}{2}A_3-hA_5\right)\right]\frac{f'}{rf}
\notag\\
&&
+A_6+\frac{1}{r^2}\left(\frac{\phi'}{2}A_2-hA_4+A_8-2A_{10}\right)\,.
\ea
In shift-symmetric Horndeski theories where the couplings 
$G_j$ $(j=2,3,4,5)$ contain the $X$ dependence alone, 
we have $\partial {\cal E}/\partial\phi=0$ 
in Eq.~(\ref{Ephi}). In this case, the scalar-field equation 
reduces to $J^r=(Q/r^2)\sqrt{h/f}$, where $Q$ is a constant.

Combining Eqs.~(\ref{back1})-(\ref{back3}), and (\ref{back4}), the 
scalar-field equation (\ref{Ephi}) can be also expressed in the form 
\be
{\cal E}_{\phi} \equiv -\frac{2}{\phi'}\left[\frac{f'}{2f}{\cal E}_{00}+{\cal E}_{11}'
+\left(\frac{f'}{2f}+\frac{2}{r}\right){\cal E}_{11}+\frac{2}{r}{\cal E}_{22}
+{\cal E}_P\right]=0\,.
\label{back5}
\ee
For a given equation of state $P=P(\rho)$ and a boundary condition 
around the center of star, we can solve Eqs.~(\ref{back1}), (\ref{back2}), (\ref{back4}), 
and (\ref{Ephi}) for $f$, $h$, $P$, and $\phi$.

\subsection{Decomposition of perturbations}

We consider metric perturbations $h_{\mu \nu}$ on top of the background metric 
$\bar{g}_{\mu \nu}$ besides perturbations of the scalar field and perfect fluid. 
In doing so, we classify them into the odd- and even-parity 
sectors depending on the types of parities under the rotation 
in two-dimensional plane $(\theta, \varphi)$ \cite{Regge:1957td,Zerilli:1970se}.
We expand the perturbations in terms of the spherical harmonics 
$Y_{lm} (\theta, \varphi)$. Under the two-dimensional rotation, 
the odd- and even-modes have the parities $(-1)^{l+1}$ and 
$(-1)^l$, respectively. Any scalar perturbation has the even-mode alone, 
whereas vector and tensor perturbations contain both odd- and 
even-modes.

To decompose the perturbations into the odd- and even-parity sectors, 
we introduce a tensor field 
\be
E_{ab}=\sqrt{\gamma}\,\varepsilon_{ab}\,,
\ee
where $\gamma$ is the determinant of two-dimensional metric given 
by the line element ${\rm d}s_{(2)}^2=\gamma_{ab}
{\rm d}x^a {\rm d} x^b
={\rm d}\theta^2+\sin^2 \theta {\rm d}\varphi^2$ 
(with $a, b$ either $\theta, \varphi$) 
and $\varepsilon_{ab}$ is an anti-symmetric symbol with 
$\varepsilon_{\theta \varphi}=1$. 
According to this definition of $E_{ab}$, we have 
$E_{\theta \varphi}=-E_{\varphi \theta}=\sin \theta$.
Any scalar perturbation $S$ as well as the $\theta, \varphi$ 
components of vector and tensor perturbations $V_{\mu}$ 
and $T_{\mu \nu}$ can be expressed, respectively, as 
\ba
S &=& \sum_{l,m} S_1 (t,r) Y_{lm} (\theta, \varphi)\,,\label{Sdec}\\
V_{a} &=& \sum_{l,m} V_1(t,r) \nabla_{a} Y_{lm}(\theta, \varphi)
+\sum_{l,m} V_2(t, r) E_{ab} \nabla^b Y_{lm}(\theta, \varphi)\,,\\
T_{a b} &=&
\sum_{l,m}\left[T_1(t,r) g_{ab}Y_{lm}
+T_2(t,r)\nabla_a\nabla_bY_{lm} \right]+
\frac{1}{2}\sum_{l,m}
T_3 (t,r) \left[
E_{a}{}^c \nabla_c\nabla_b Y_{lm}
+ E_{b}{}^c \nabla_c\nabla_a Y_{lm}
\right]\,,\label{Tdec}
\ea
where $S_1$, $V_1$, $V_2$, $T_1$, $T_2$, and $T_3$ are 
functions of $t$ and $r$ (for which we omit the subscripts 
``$lm$''), and $\nabla_a$ is 
a covariant derivative operator defined on the two-dimensional 
sphere. The odd- and even-parity modes correspond to 
those containing the functions $V_2$, $T_3$ and 
$S_1$, $V_1$, $T_1$, $T_2$, 
respectively \cite{DeFelice:2011ka,Motohashi:2011pw,Kase:2014baa}.

In Secs.~\ref{oddsec} and \ref{evensec}, we first apply the decompositions 
(\ref{Sdec})-(\ref{Tdec}) to the perturbations associated with fundamental fields 
$h_{\mu \nu}$, $\phi$, as well as the scalar and vector quantities, 
$\ell$, $J^{\mu}$, ${\cal A}_i$, and ${\cal B}_i$, appearing in the matter 
action (\ref{SM}). Then, we expand all the quantities composed of these fundamental fields 
in the action (\ref{action}), e.g., Ricci scalar $R$, kinetic term $X$ of the scalar field, 
number density $n$, energy density $\rho$, up to second order in perturbations. 
The resulting quadratic-order actions of odd- and even-parity sectors will be used 
to derive the perturbation equations of motion and stability conditions.

%%%%%%%%%%%%%%%%%%%%%%%%%%%%%%%%%%%%%%%%%%%
\section{Odd-parity perturbations}
\label{oddsec}
%%%%%%%%%%%%%%%%%%%%%%%%%%%%%%%%%%%%%%%%%%%

In this section, we study the propagation of odd-parity perturbations 
in Horndeski theories with the matter action given by (\ref{SM}). 
Under a rotation in the $(\theta, \varphi)$ plane, the $h_{tt}$, $h_{tr}$, 
and $h_{rr}$ components of metric perturbations $h_{\mu \nu}$ transform 
as scalars. The $h_{ta}$ and $h_{ra}$ components 
behave as vectors (with $a, b$ either $\theta, \varphi$), 
while $h_{ab}$ transforms as a tensor. 
Then, from Eqs.~(\ref{Sdec})-(\ref{Tdec}), the metric components 
in the odd-parity sector 
are given by 
\ba
& &
h_{tt}=h_{tr}=h_{rr}=0\,,\\
& &
h_{ta}=\sum_{l,m}Q(t,r)E_{ab}
\nabla^bY_{lm}(\theta,\varphi)\,,
\qquad
h_{ra}=\sum_{l,m}W(t,r)E_{ab} \nabla^bY_{lm}(\theta,\varphi)\,,\\
& &
h_{ab}=
\frac{1}{2}\sum_{l,m}
U (t,r) \left[
E_{a}{}^c \nabla_c\nabla_b Y_{lm}(\theta,\varphi)
+ E_{b}{}^c \nabla_c\nabla_a Y_{lm}(\theta,\varphi)
\right]\,,
\ea
where $Q$, $W$, and $U$ are functions of $t$ and $r$. 
The scalar perturbation in the odd-parity sector vanishes, 
so the field $\phi$ has the background value 
$\bar{\phi}(r)$ alone, i.e., 
\be
\phi=\bar{\phi}(r)\,.
\ee
In the following, we omit a bar for brevity. 

For the matter sector, the scalar quantities $\rho$ and $n$ 
appearing in the action (\ref{SM}) do not have odd-parity 
perturbations. The components of $J_{\mu}$ 
associated with the odd-parity sector are given by 
\be
J_t=\bar{J}_t\,,\qquad
J_{r}=0\,,\qquad 
J_{a}=\sum_{l,m} \sqrt{-\bar{g}}\,\delta j(t,r)E_{ab}
\nabla^bY_{lm}(\theta,\varphi)\,,
\label{Jodd}
\ee
where $\bar{J}_t=-n(f/\sqrt{h})r^2 \sin \theta$
and $\sqrt{-\bar{g}}=\sqrt{f/h}\,r^2 \sin \theta$ are
the background values. 
{}From Eq.~(\ref{defu}), the perturbation $\delta j$ 
is related to the $\theta$, $\varphi$ components of four 
velocity $u^a$.
The spatial vectors ${\cal A}_i$ and ${\cal B}^i$, 
where $i=r, \theta, \varphi$,  are 
expressed in the forms 
\ba 
{\cal A}_i &=& 
\delta {\cal A}_i\,,\qquad \qquad {\rm with} \quad
\delta {\cal A}_r=0\,,\qquad 
\delta {\cal A}_a=\sum_{l,m}\delta {\cal A}(t,r)
E_{ab} \nabla^bY_{lm}(\theta,\varphi)\,,\\
{\cal B}^i &=& 
x^i+\delta {\cal B}^i\,,\qquad {\rm with} \quad
\delta {\cal B}^r=0\,,\qquad 
\delta {\cal B}^a=\sum_{l,m}\delta {\cal B}(t,r)
{E^{a}}_{b} \nabla^bY_{lm}(\theta,\varphi)\,.
\label{cBi}
\ea
The perturbations $\delta {\cal A}$ and $\delta {\cal B}$ are related to 
$\delta j$ according to Eqs.~(\ref{JAB1}) and (\ref{JAB2}).
{}From Eq.~(\ref{rhomu}), the Lagrange multiplier $\ell$ in the odd-parity sector 
obeys $\partial_t \ell=\rho_{,n}(r) u_t$ with $u_t=-\sqrt{f(r)}$. 
Integrating this relation gives
\be
\ell=-\rho_{,n}(r) \sqrt{f(r)}\,t\,,
\label{ellt}
\ee
where the integration constant is set to 0 (which does not affect 
the discussion below).

In what follows, we derive the second-order action of odd-parity 
perturbations for the purpose of obtaining linear perturbation 
equations of motion and associated stability conditions. 
Due to the spherical symmetry of the background, we can restrict 
ourselves to axisymmetric modes of perturbations ($m=0$) by 
the suitable rotation of nonaxisymmetric modes 
($m \neq 0$) \cite{deRham:2020ejn}.
Hence we will set $m=0$ without loss of generality and 
multiply $2\pi$ for the integral with respect to $\varphi$.

\subsection{Second-order action}

We expand the action (\ref{action}) up to second order in odd-parity perturbations 
for the multipoles $l \geq 2$ without choosing particular gauge conditions. 
The perturbation $\delta \phi$ of the scalar field $\phi$ vanishes, but 
the kinetic energy $X$ acquires a second-order contribution 
$\delta X=-h^2 \phi'^2 W^2 Y_{l0, \theta}^2/(2r^2)$ besides its 
background value $\bar{X}=-h \phi'^2/2$. 
{}From Eq.~(\ref{defn}), the perturbation of the fluid number 
density expanded up to second-order is given by 
\be
\delta n=\frac{n}{2r^2}\left(hW^2-\frac{2}{f}Q^2
+\frac{2}{n\sqrt{f}}Q\delta j-\frac{\delta j^2}{n^2}
+\frac{U^2}{4r^2} \cot^2 \theta \right)
Y_{l0, \theta}^2
-\frac{nU^2}{4r^4} \left( Y_{l0, \theta} \cot \theta 
-\frac{1}{2} Y_{l0, \theta \theta} \right)Y_{l0, \theta \theta}\,,
\ee
where $n$ is the background value. 
The matter density in (\ref{SM}) contains its background value $\rho(r)$ 
and the perturbation $\rho_{,n} \delta n=(\rho+P)\delta n/n$.
We also exploit Eqs.~(\ref{Jodd})-(\ref{cBi}) for the expansion of 
the Schutz-Sorkin action.

After expanding the total action (\ref{action}) up to second order in perturbations, 
there are $\theta$-dependent terms containing the products such as 
$Y_{l0, \theta}Y_{l0, \theta \theta \theta}$, 
$Y_{l0, \theta}Y_{l0, \theta \theta}$, etc. 
After the integrations by parts, the resulting action possesses the terms 
proportional to $Y_{l0, \theta}^2$ and $Y_{l0, \theta \theta}^2$. 
These $\theta$-dependent terms can be integrated out from the action 
by employing the following relations \cite{Kase:2018voo}
\be
\int_0^{2\pi}{\rm d} \varphi \int_0^{\pi} {\rm d} \theta\,
Y_{l0, \theta}^2 \sin \theta=L\,,\qquad
\int_0^{2\pi}{\rm d} \varphi \int_0^{\pi} {\rm d} \theta\, 
\left( \frac{Y_{l0, \theta}^2}{\sin \theta}+
Y_{l0, \theta \theta}^2 \sin \theta \right)=L^2\,,
\label{Ythere}
\ee
where 
\be
L \equiv l (l+1)\,.
\ee
After this procedure, we perform the integrations by parts 
with respect to $t$ and $r$, and exploit the background 
equations presented in Sec.~\ref{backsec}.
Then, the second-order action in the odd-parity sector yields 
\be
{\cal S}_{\rm odd}=\sum_{l} L \int {\rm d}t {\rm d}r\,{\cal L}_{\rm odd}\,,
\label{Sodd}
\ee
where 
\ba
{\cal L}_{\rm odd} &=& \frac{\sqrt{h}}{4\sqrt{f}} {\cal H} 
\left( \dot{W}-Q'+\frac{2Q}{r} \right)^2
+(L-2) \left( \frac{{\cal F}Q^2}{4r^2 \sqrt{fh}}
-\frac{\sqrt{fh}}{4r^2} {\cal G}W^2 \right)
+{\cal L}_U\nonumber \\
&&+\frac{\sqrt{f}(\rho+P)}{2n^2 \sqrt{h}} \delta j^2
-\frac{n (r^2\dot{\delta {\cal B}}-Q)+\sqrt{f} 
\delta j}{\sqrt{h}} \delta {\cal A}\,,
\label{Lodd}
\ea
with a dot being the derivative with respect to $t$. 
The variables ${\cal H}$, ${\cal F}$, and ${\cal G}$
are defined by 
\ba
{\cal H}&\equiv&2 G_4+2 h\phi'^2G_{4,X}-h\phi'^2G_{5,\phi}
-\frac{h^2 \phi'^3 G_{5,X}}{r}
\,,\label{cHdef}\\
{\cal F}&\equiv&2 G_4+h\phi'^2G_{5,\phi}-h\phi'^2  \left( \frac12 h' \phi'+h \phi'' \right) G_{5,X}\,,
\label{cFdef}\\
{\cal G}&\equiv&2 G_4+2 h\phi'^2G_{4,X}-h\phi'^2 \left( G_{5,\phi}+{\frac {f' h\phi' G_{5,X}}{2f}} \right) \,.
\label{cGdef}
\ea
The Lagrangian ${\cal L}_U$, which arises from the gravitational perturbation $U$, 
is given by 
\ba
{\cal L}_U &=&\frac{L-2}{8r^2 \sqrt{fh}} 
\biggl[ \{ 2 {\cal F} \dot{Q}-2fh {\cal G}W'
-(f' h {\cal G}+fh' {\cal G}+2fh {\cal G}')W \} U
+\frac{1}{2} {\cal F} \dot{U}^2-\frac{1}{2}fh{\cal G}U'^2 \nonumber \\
& &\qquad \qquad~-\frac{1}{2r^2}
(rf' h {\cal G}+rfh' {\cal G}+2rfh {\cal G}'-2fh {\cal G})U^2
\biggl]\,.
\ea
In the absence of the perfect fluid with the particular gauge choice $U=0$, 
the action (\ref{Sodd}) coincides with that derived in Ref.~\cite{Kobayashi:2012kh}.

\subsection{Perturbation equations}
\label{peroddsec}

The second-order action (\ref{Sodd}) is written in a gauge-ready 
form\footnote{On an isotropic and homogenous cosmological 
background, the second-order perturbed action of full Horndeski 
theories was also derived in a gauge-ready form in 
Ref.~\cite{Kase:2018aps}.}, so 
we can derive the linear perturbation equations of motion for any 
gauge choices of interest.
Under the gauge transformation $x_{\mu} \to x_{\mu}+\xi_{\mu}$, 
where $\xi_t=0$, $\xi_r=0$, and 
$\xi_a=\sum_{l} \Lambda(t, r) E_{ab} \nabla^b Y_{lm} (\theta, \varphi)$, 
the three gravitational perturbations $Q$, $W$, and $U$ 
transform, respectively, as
\be
Q \to Q+\dot{\Lambda}\,,\qquad 
W \to W+\Lambda'-\frac{2}{r}\Lambda\,,\qquad
U \to U+2\Lambda\,.
\label{gaugeodd}
\ee
A commonly used gauge-invariant variable is 
\be
\chi \equiv \dot{W}-Q'+\frac{2Q}{r}\,,
\ee
which was first introduced by Cunningham {\it et al.} \cite{Cunningham:1978zfa}.
This corresponds to a dynamical perturbation arising from 
the gravity sector. 

First of all, varying the Lagrangian (\ref{Lodd}) with respect to $U$ leads to 
\ba
& &
r^2 \left[ 2 {\cal F} \dot{Q}-2fh {\cal G}W'
-(f' h {\cal G}+fh' {\cal G}+2fh {\cal G}')W \right]
-r^2{\cal F} \ddot{U}+r^2fh{\cal G}U''+r^4 \sqrt{fh} \left( 
\frac{\sqrt{fh} {\cal G}}{r^2}\right)'U'  \nonumber \\
& &
-(rf' h {\cal G}+rfh' {\cal G}+2rfh {\cal G}'-2fh {\cal G})U=0\,.
\label{QtWW}
\ea
In the following, we choose the Regge-Wheeler gauge \cite{Regge:1957td}
\be
U=0\,,
\ee
under which the gauge transformation scalar $\Lambda$ in 
Eq.~(\ref{gaugeodd}) is completely fixed. 
Then, Eq.~(\ref{QtWW}) reduces to 
\be
2 {\cal F} \dot{Q}-2fh {\cal G}W'
-(f' h {\cal G}+fh' {\cal G}+2fh {\cal G}')W =0\,.
\label{QtWW2}
\ee
As we will see below, this relation is consistent with the 
other perturbation equations of motion

To derive the perturbation equation of $\chi$, we first vary the Lagrangian
(\ref{Lodd}) with respect to $W$ and express $\ddot{W}$ 
by using $\dot{\chi}$. This leads to 
\be
W =-\frac{r^2{\cal H}}{(L-2) f{\cal G}} \dot{\chi}\,.
\label{Wre}
\ee
Taking the time derivative of this relation and replacing $\dot{W}$
with $\chi+Q'-2Q/r$, it follows that 
\be
r^2 {\cal H} \ddot{\chi}+\left( L-2 \right) f{\cal G} 
\left( \chi+Q'-\frac{2Q}{r} \right)=0\,.
\label{ddtotchieq}
\ee
Varying (\ref{Lodd}) with respect to $Q$, we obtain 
\be
Q=\frac{r}{2(L-2)f {\cal F}} \left[(rhf'-rf h'-4fh) {\cal H} \chi 
-2rfh ({\cal H}' \chi+{\cal H}\chi')-4r f^{3/2}n \delta {\cal A}
\right]\,.
\label{Qre}
\ee
We take the $r$ derivative of Eq.~(\ref{Qre}) and express 
$Q'$ and $Q$ in terms of $\chi$, 
$\delta {\cal A}$, and their $r$ derivatives. 
Then, Eq.~(\ref{ddtotchieq}) gives 
\be
\ddot{\chi}-fh \frac{{\cal G}}{{\cal F}} \chi''
+\alpha_1 \chi'
+\frac{f}{r^2} \frac{{\cal G}}{{\cal H}}\left( L-2
+\alpha_2'-\frac{2}{r} \alpha_2 \right)\chi
-\frac{\sqrt{f} {\cal G}}{{\cal F}^2{\cal H}} 
\left[ (n f' {\cal F}+2n'f {\cal F}-2n f{\cal F}') \delta {\cal A}
+2n f {\cal F} \delta {\cal A}' \right]=0\,,
\label{chi0}
\ee
where 
\ba
\alpha_1 &\equiv& 
-\frac{{\cal G}}{2r {\cal F}^2 {\cal H}} \left[ \{ 
(4fh+3rfh'-rf'h){\cal H}+4rfh {\cal H}' \} {\cal F} 
-2rfh{\cal H} {\cal F}' \right]\,,\\
\alpha_2 &\equiv& 
-\frac{r^2h{\cal H}}{\cal F}\left(\frac{{\cal H}'}{\cal H}-\frac{f'}{2f}
+\frac{h'}{2h}+\frac{2}{r}\right)
\,.
\ea

The gravitational perturbation $\chi$ is coupled to the perfect fluid 
through the perturbation $\delta {\cal A}$. 
Varying the Lagrangian (\ref{Lodd}) with respect to $\delta {\cal A}$, 
$\delta {\cal B}$, and $\delta j$, respectively, we obtain
\ba
& &
\dot{\delta {\cal B}}=\frac{1}{r^2} \left( Q-\frac{\sqrt{f}}{n} 
\delta j \right)\,,\label{oddma1}\\
& &
\dot{\delta {\cal A}}=0\,,\label{oddma2}\\
& &
\delta {\cal A}=\frac{\rho+P}{n^2} \delta j\,.
\label{oddma3}
\ea
Taking the $t$ and $r$ derivatives of Eq.~(\ref{Qre}) and Eq.~(\ref{Wre}), 
respectively, and using Eq.~(\ref{oddma2}), one can confirm that 
Eq.~(\ref{QtWW2}) is consistently satisfied.
{}From Eq.~(\ref{oddma3}), the Lagrange multiplier $\delta {\cal A}$ is related to 
the physical perturbation $\delta j$. 
As we see in Eqs.~(\ref{defu}) and (\ref{Jodd}), $\delta j$ corresponds to 
components $u_a$ of the fluid four velocity. 
{}From Eq.~(\ref{oddma2}), $\delta {\cal A}$ depends on $r$ alone, 
so that 
\be
\dot{\delta j}=0\,.
\label{deltaj}
\ee
Substituting Eq.~(\ref{oddma3}) and its $r$ derivative 
into Eq.~(\ref{chi0}), we find
\be
\ddot{\chi}-fh \frac{{\cal G}}{{\cal F}} \chi''
+\alpha_1 \chi'
+\frac{f}{r^2} \frac{{\cal G}}{{\cal H}}\left( L-2
+\alpha_2'-\frac{2}{r} \alpha_2 \right)\chi
=\frac{2f^{3/2}(\rho+P){\cal G}}{n {\cal F}^2 {\cal H}} 
\left( {\cal F} \delta j'-{\cal F}' \delta j \right)\,,
\label{chieq}
\ee
so that the field $\chi$ is coupled to the perturbation $\delta j$.
Since $\delta j$ is nondynamical due to the relation (\ref{deltaj}), 
the gravitational perturbation $\chi$ is the only dynamical degree 
of freedom. We have thus shown that the dynamics of odd-parity 
perturbation is governed by Eq.~(\ref{chieq}) with the nondynamical 
perturbation satisfying Eq.~(\ref{deltaj}). 
These results are consistent with those derived in GR with a perfect fluid,  
see Eqs.~(II-12a) and (II-12b) in Ref.~\cite{Cunningham:1978zfa}.

Assuming the solution of Eq.~(\ref{chieq}) in the form 
\be
\chi=\chi_0 e^{i (\omega t-k r-l \theta)}\,,
\ee
where $\chi_0$, $\omega$, $k$, $l$ are constants, we derive 
the dispersion relation and propagation speeds of $\chi$.
In the limits of large $\omega$ and $k$, and $l \gg 1$, we can ignore the 
terms $\alpha_1 \chi'$, $\alpha_2' \chi$, $\alpha_2 \chi/r$, 
and the right hand-side of Eq.~(\ref{chieq}), and hence
\be
\omega^2-fh \frac{{\cal G}}{{\cal F}} k^2 -l^2
\frac{f}{r^2} \frac{{\cal G}}{{\cal H}}=0\,.
\label{dispe}
\ee
In terms of the proper time $\tau=\int \sqrt{f}{\rm d}t$ and the rescaled 
radial coordinate $r_*=\int {\rm d}r/\sqrt{h}$, the radial propagation 
speed is $c_r={\rm d}r_*/{\rm d}\tau=\hat{c}_r/\sqrt{fh}$, where 
$\hat{c}_r={\rm d}r/{\rm d}t=\omega/k$. 
To derive $c_r$, we drop the third term in Eq.~(\ref{dispe}) 
and substitute $\omega=k\sqrt{fh}\,c_r$ into Eq.~(\ref{dispe}).
This leads to
\be
c_r^2=\frac{{\cal G}}{{\cal F}}\,.
\label{crodd}
\ee
Similarly, the angular propagation speed in proper time is 
$c_{\Omega}=r {\rm d} \theta/{\rm d}\tau=\hat{c}_{\Omega}/\sqrt{f}$, 
where $\hat{c}_{\Omega}=r {\rm d} \theta/{\rm d}t$ satisfies
$\omega^2=\hat{c}_{\Omega}^2 l^2/r^2$. 
Substituting $\omega^2=c_{\Omega}^2fl^2/r^2$ into Eq.~(\ref{dispe}) 
and dropping the second term in Eq.~(\ref{dispe}), we obtain
\be
c_\Omega^2=\frac{{\cal G}}{{\cal H}}\,.
\label{cOodd}
\ee
To avoid the Laplacian instabilities along the radial and angular 
directions, we require the two conditions $c_r^2 \geq 0$ 
and $c_\Omega^2 \geq 0$. These conditions coincide 
with those derived in Ref.~\cite{Kobayashi:2012kh} in the absence of matter.

In Ref.~\cite{Kase:2020qvz}, the relation (\ref{oddma1}) is plugged into 
the action (\ref{Lodd}), after which the term proportional to $\delta {\cal A}$ 
vanishes from the action. 
Then, we are left with the matter action proportional to $\delta j^2$.
Variation of this action with respect to $\delta j$ leads to $\delta j=0$, 
in which case all the matter perturbations in (\ref{Lodd}) vanish.
However, this procedure is not correct. The perturbation 
$\delta {\cal A}$ appears in the action only as a linear form 
$f(\delta j, Q, \dot{\delta {\cal B}}) \delta {\cal A}$ and hence 
the variation with respect to $\delta {\cal A}$ leads to $f=0$, i.e., Eq.~(\ref{oddma1}).
The incorrect procedure is to substitute $f=0$ into 
the Lagrangian (\ref{Lodd}), after which the product $\delta j \delta {\cal A}$ 
disappears. When we vary the original Lagrangian (\ref{Lodd}) 
with respect to $\delta j$, the presence of this product gives rise 
to the coupling between $\delta {\cal A}$ and $\delta j$, 
see Eq.~(\ref{oddma3}).
Indeed, we have shown that the 
perturbation $\chi$ is coupled to $\delta j$ through the relation 
(\ref{oddma3}). 
In Ref.~\cite{Kase:2020qvz}, the perturbation $\chi$ was 
introduced as a Lagrange multiplier at the action level,
so we also take this approach in Sec.~\ref{acsec} to clarify the 
issue of the elimination of nondynamical variables from the action.

\subsection{Action approach}
\label{acsec}

For the gauge choice $U=0$, the second-order Lagrangian (\ref{Lodd}) is 
equivalent to 
\ba
{\cal L}_{\rm odd} &=& \frac{\sqrt{h}}{4\sqrt{f}} {\cal H} 
\left[ 2\chi \left( \dot{W}-Q'+\frac{2Q}{r} \right)-\chi^2 
\right]
+(L-2) \left( \frac{{\cal F}Q^2}{4r^2 \sqrt{fh}}
-\frac{\sqrt{fh}}{4r^2} {\cal G}W^2 \right)
\nonumber \\
&&+\frac{\sqrt{f}(\rho+P)}{2n^2 \sqrt{h}} \delta j^2
-\frac{n (r^2\dot{\delta {\cal B}}-Q)+\sqrt{f} 
\delta j}{\sqrt{h}} \delta {\cal A}\,, 
\label{Lodd2}
\ea
whose variation with respect to $\chi$ gives 
$\chi=\dot{W}-Q'+2Q/r$.
Varying (\ref{Lodd2}) with respect to $W$ and $Q$, 
respectively, we obtain the same equations as 
Eqs.~(\ref{Wre}) and (\ref{Qre}).
The Lagrangian (\ref{Lodd2}) contains the terms $W^2$ and $Q^2$ besides 
$\dot{W}$, $Q'$ and $Q$ (i.e., they do not appear as 
linear terms), so we can substitute the relations 
(\ref{Wre}) and (\ref{Qre}) as well as their $t$ and $r$ 
derivatives into Eq.~(\ref{Lodd2}).
After the integrations by parts, the action reduces to 
the form (\ref{Sodd}), with the Lagrangian 
\be
{\cal L}_{\rm odd}=K_{\chi} \dot{\chi}^2+G_{\chi} \chi'^2
+M_{\chi} \chi^2+{\cal L}_m\,,
\label{Loddf}
\ee
where 
\ba
& &
K_{\chi}=\frac{r^2 \sqrt{h}{\cal H}^2}{4(L-2)f^{3/2}{\cal G}}\,,
\qquad
G_{\chi}=-fh \frac{{\cal G}}{{\cal F}}K_{\chi}\,,
\qquad
M_{\chi}=-K_{\chi}
\frac{f}{r^2} \frac{{\cal G}}{{\cal H}}\left( L-2
+\alpha_2'-\frac{2}{r} \alpha_2 \right)
\,,\\
& &
{\cal L}_m=\frac{1}{2(L-2)f^{3/2}\sqrt{h} n^2 {\cal F}} 
[ (L-2) (\rho+P) f^2 {\cal F} \delta j^2
-2(L-2)f^2 n^2 {\cal F} \delta j \delta {\cal A}
-2r^2 f^2 n^4 \delta {\cal A}^2 \nonumber \\
& &\qquad\quad 
-\{ (2r (L-2){\cal F} \dot{\delta {\cal B}}+r (h' {\cal H}+2h {\cal H}') \chi
+2rh {\cal H}\chi'+4h {\cal H}\chi)rf^{3/2}n
-\sqrt{f}f'h n r^2 {\cal H} \chi \} n^2 \delta {\cal A} ]\,.
\label{Lm}
\ea
The matter Lagrangian ${\cal L}_m$ contains terms $\delta j^2$ 
and $\delta j \delta {\cal A}$, so the variation with respect to $\delta j$ 
gives $\delta {\cal A}=(\rho+P)\delta j/n^2$, i.e., Eq.~(\ref{oddma3}). 
Substituting this relation into Eq.~(\ref{Lm}) to eliminate the Lagrange 
multiplier $\delta {\cal A}$, we obtain the reduced matter Lagrangian
\ba
{\cal L}_m
&=&-\frac{r(\rho+P)}{n\sqrt{h}} \bigg[ 
r \delta j \dot{\delta {\cal B}}
+\frac{\sqrt{f}\{ (L-2){\cal F}+2r^2 (\rho+P)\}}{2(L-2) rn {\cal F}} 
\delta j^2\nonumber \\
&&
+\frac{rh {\cal H}}{(L-2){\cal F}} \chi' \delta j  
+\frac{\{r(h' {\cal H}+2h {\cal H}')+4h {\cal H}\}f-rf'h {\cal H}}{2(L-2)f {\cal F}}
\chi \delta j \biggr]\,.
\label{Loddre}
\ea
Varying Eq.~(\ref{Loddre}) with respect to $\delta {\cal B}$, 
the product $\delta j \dot{\delta {\cal B}}$ gives the relation 
$\dot{\delta j}=0$, i.e., Eq.~(\ref{deltaj}). 
Hence the perturbation $\delta j$ is nondynamical. 
Varying Eq.~(\ref{Loddre}) with respect to $\delta j$ and 
using Eq.~(\ref{Qre}), we obtain the same equation as (\ref{oddma1}). 
The dynamical perturbation $\chi$ is coupled to $\delta j$ through 
the last two terms in Eq.~(\ref{Loddre}). 
Indeed, varying Eq.~(\ref{Loddf}) with the reduced matter 
Lagrangian (\ref{Loddre}) in terms of $\chi$ leads to the same equation 
as (\ref{chieq}). This shows the consistency of the action approach 
taken above.

Since the field $\chi$ is coupled to $\delta j$ through the 
Lagrangians $\chi' \delta j$ and $\chi \delta j$, the interaction with 
$\delta j$ does not affect the propagation of $\chi$ in the limits of 
large $\omega$, $k$, and $l$. 
In this limit, the Lagrangian (\ref{Loddf}) shows that 
the ghost is absent under the condition $K_{\chi}>0$. 
This translates to the condition
\be
{\cal G}>0\,.
\label{nogoodd}
\ee
In the approach taken in Sec.~\ref{peroddsec}, we derived the perturbation 
equations from the original action (\ref{Sodd}), in which case the no-ghost 
condition (\ref{nogoodd}) was not explicit. 
In the limits of large $\omega$, $k$, and $l$, the reduced Lagrangian 
(\ref{Loddf}) leads to the same dispersion relation as Eq.~(\ref{dispe}). 
Hence the gravitational perturbation $\chi$ has the same radial and angular 
propagations speeds as those given in Eqs.~(\ref{crodd}) and (\ref{cOodd}), 
respectively. Combining the no-ghost condition (\ref{nogoodd}) with 
the Laplacian stability conditions $c_r^2 \geq 0$ and $c_\Omega^2 \geq 0$,  
we require that 
\be
{\cal F}>0\,,\qquad {\cal H}>0\,, 
\label{FH}
\ee
which coincide with those derived in Ref.~\cite{Kobayashi:2012kh} in the absence 
of the perfect fluid.
The results given above are valid for $l \geq 2$, but the situation is different 
for $l=1$. Since $L=2$ for $l=1$, the terms proportional to $(L-2)$ vanish from 
Eq.~(\ref{Lodd2}). Then, variation of the Lagrangian (\ref{Lodd2}) with respect to 
$W$ gives $\dot{\chi}=0$. Since the property $\dot{\delta j}=0$ also holds, 
both $\chi$ and $\delta j$ do not propagate for $l=1$.

%%%%%%%%%%%%%%%%%%%%%%%%%%%%%%%%%%%%%%%%%%%
\section{Even-parity perturbations}
\label{evensec}
%%%%%%%%%%%%%%%%%%%%%%%%%%%%%%%%%%%%%%%%%%%

Let us proceed to the derivation of the second-order action and perturbation 
equations of motion for even-parity modes.
The nonvanishing components of metric perturbations in 
the even-parity sector are given by\footnote{
In Ref.~\cite{Kase:2020qvz}, the notation $\alpha$ 
was used for $h_1$.} \cite{Regge:1957td}
\ba
&&
h_{tt}=f(r) \sum_{l,m}H_0(t,r)Y_{lm}(\theta,\varphi)\,,\qquad
h_{tr}=h_{rt}=\sum_{l,m}H_1(t,r)Y_{lm}(\theta,\varphi)\,,\qquad
h_{rr}=h(r)^{-1}\,\sum_{l,m}H_2(t,r)Y_{lm}(\theta,\varphi)\,,\notag\\
&&
h_{ta}=h_{at}=\sum_{l,m}h_0(t,r)\nabla_aY_{lm}(\theta,\varphi)\,,\qquad
h_{ra}=h_{ar}=\sum_{l,m}h_1(t,r)\nabla_aY_{lm}(\theta,\varphi)\,,\notag\\
&&
h_{ab}=\sum_{l,m}\left[K(t,r)g_{ab}Y_{lm}(\theta,\varphi)
+r^2G(t,r)\nabla_a\nabla_bY_{lm}(\theta,\varphi)\right]\,, 
\label{meteven}
\ea
where $H_0$, $H_1$, $H_2$, $h_0$, $h_1$, $K$, and $G$ 
are functions of $t$ and $r$. 
The scalar field $\phi$ is decomposed into the background 
value $\bar{\phi}(r)$ and the perturbation $\delta\phi (t, r)$ 
as 
\be
\phi=\bar{\phi}(r)+\sum_{l,m} \delta \phi(t,r) 
Y_{lm} (\theta,\varphi)\,.
\label{phidec}
\ee
For the matter sector, the components of $J_{\mu}$ containing 
even-parity perturbations are expressed 
in the forms \cite{Kase:2020qvz}
\be
J_{t}=\bar{J}_t+\sum_{l,m} \sqrt{-\bar{g}}\,\delta J_t (t,r)Y_{lm}(\theta,\varphi)\,,\qquad
J_{r}=\sum_{l,m} \sqrt{-\bar{g}}\,\delta J_r (t,r)Y_{lm}(\theta,\varphi)\,,\qquad
J_{a}=\sum_{l,m} \sqrt{-\bar{g}}\,\delta J(t,r) \nabla_aY_{lm}(\theta,\varphi)\,.
\label{Jeven}
\ee
We will mostly omit a bar from the background quantities. 
The intrinsic spatial vector fields ${\cal A}_i$ and ${\cal B}^i$ 
are given by 
\be
{\cal A}_i=\delta {\cal A}_i\,,\qquad 
{\cal B}^i=x^i+\delta {\cal B}^i\,,
\label{ABi}
\ee
with the perturbed components,
\ba
& &
\delta {\cal A}_r=\sum_{l,m}\delta {\cal A}_1(t,r)Y_{lm}(\theta,\varphi)\,,\qquad 
\delta {\cal A}_a=\sum_{l,m}\delta {\cal A}_2(t,r)\nabla_aY_{lm}(\theta,\varphi)\,,
\label{delAa}\\
&&
\delta {\cal B}^r=\sum_{l,m}\delta {\cal B}_1(t,r)Y_{lm}(\theta,\varphi)\,,\qquad 
\delta {\cal B}^a=\sum_{l,m}\delta {\cal B}_2(t,r)\nabla_aY_{lm}(\theta,\varphi)\,.
\label{delBa}
\ea
Unlike Ref.~\cite{Kase:2020qvz}, we define $\delta {\cal B}_1$ and $\delta {\cal B}_2$ as 
the components of $\delta {\cal B}^{i}$ (i.e., upper index).
The Lagrangian multiplier $\ell$ is decomposed as 
\be
\ell=-\rho_{,n} (r) \sqrt{f(r)}\,t+\sum_{l,m} \delta\ell(t,r) Y_{lm} (\theta, \varphi)\,,
\label{ldef}
\ee
where the first term on the right hand-side is the background 
value, see Eq.~(\ref{ellt}). 
The explicit form of $\delta\ell$ is known from Eq.~(\ref{rhomu}). 
Similarly, it is also determined by using the perturbation equation of 
$J_{\mu}$ derived later. In the following, we will take the latter approach.

The matter density perturbation $\delta \rho (t,r)$ is related to 
the first-order number density perturbation $\delta n$, as
\be
\delta n=\sum_{l,m} \frac{\delta \rho (t,r)}{\rho_{,n}(r)} 
Y_{lm} (\theta, \varphi)\,.
\label{delnrho}
\ee
To compute the second-order action of even-parity perturbations, 
we set $m=0$ without loss of generality.
Expanding Eq.~(\ref{defn}) on account of Eqs.~(\ref{meteven}) 
and (\ref{Jeven}), the first-order perturbation of $n$ 
is expressed in the form
\be
\delta n=-\sum_{l}\left[\frac{\delta J_t}{\sqrt{f}}
+n\left(\frac{1}{2}H_2-H_0+K-\frac{L}{2}G\right)\right]
Y_{l0} (\theta, \varphi)+{\cal O}(\varepsilon^2)\,,
\label{deln}
\ee
where $\varepsilon^n$ represents the $n$-th order of perturbations, 
and  we used the property
\be
Y_{l0, \theta \theta}+(\cot \theta) Y_{l0, \theta}
+L Y_{l0}=0\,,
\ee
with $L=l(l+1)$. 
Comparing the first-order term of Eq.~(\ref{deln}) with Eq.~(\ref{delnrho}), 
it follows that 
\be
\delta J_t=-\sqrt{f}\left[\frac{\delta\rho}{\rho_{,n}}
+n\left(\frac{1}{2}H_2-H_0+K-\frac{L}{2}G\right)\right]\,.
\ee
We will exploit this relation to eliminate  
$\delta J_t$ from the action. 
The matter density $\rho(n)$ in Eq.~(\ref{SM}) is expanded as 
\be
\rho(n)=\bar{\rho}(n)+\rho_{,n} \delta n
+\frac{c_m^2 \rho_{,n}}{2n} \delta n^2
+{\cal O}(\varepsilon^3)\,,
\label{rhon}
\ee
where $c_m^2$ is the matter propagation speed squared defined by
\be
c_m^2 \equiv \frac{n \rho_{,nn}}{\rho_{,n}}\,.
\label{cm}
\ee
In Eq.~(\ref{deln}), we will take the second-order contribution to 
$\delta n$ into account for the computation 
of $\rho_{,n} \delta n$.

The other components of $J_{\mu}$, i.e., $\delta J_r$ and $\delta J$, can be 
expressed in terms of the perturbed components of 
fluid four velocity $u_{\mu}$. 
We write the radial and angular components of $u_{\mu}$ 
in the forms
\be
u_r=\sum_{l,m}\delta u_r (t, r) Y_{lm} (\theta, \varphi)\,, \qquad
u_a=\sum_{l,m}v (t, r) \nabla_aY_{lm} (\theta, \varphi)\,,
\ee
where $\delta u_r$ and $v$ are functions of $t$ and $r$.
Since the relation between $u_{\mu}$ and 
$J_{\mu}$ is $u_{\mu}=J_{\mu}/(n \sqrt{-g})$,  
we have
\be
\delta J_r=n\delta u_r\,,\qquad
\delta J=nv\,.
\ee
We will use the perturbations $\delta u_r$ and $v$ 
instead of $\delta J_r$ and $\delta J$ in the following discussion.

\subsection{Second-order action}
\label{secA}

We expand the action (\ref{action}) up to second order 
in even-parity perturbations without choosing 
any particular gauge conditions.
The functions $G_{j}$ ($j=2,3,4,5$) are expressed as
\be
G_j(\phi,X) = 
G_j(r)+G_{j,\phi}\delta \phi+G_{j,X}\delta X
+\frac{1}{2} G_{j,\phi \phi} \delta \phi^2
+\frac{1}{2} G_{j,XX} \delta X^2
+G_{j,\phi X} \delta \phi \delta X+{\cal O}(\varepsilon^3)\,,
\ee
where $\delta X$ is given by 
\ba
\delta X &=&
\sum_{l} \frac12 h \phi' \left( \phi' H_2- 2\delta \phi' 
\right) Y_{l0}
+\sum_{l} \frac{1}{2f} \left[ \dot{\delta \phi}^2+h^2 \phi'^2 H_1^2
-h \left\{ f\delta \phi'^2+2\phi' (H_1 \dot{\delta \phi}-f H_2 \delta \phi' )
+f \phi'^2 H_2^2 \right\} \right] Y_{l0}^2 \nonumber\\
&&-\sum_{l}\frac{1}{2r^2} \left( \delta \phi-h \phi' h_1 \right)^2 
Y_{l0, \theta}^2+{\cal O}(\varepsilon^3)\,.
\ea
After expanding the action, we can integrate out the $\theta$-dependent 
terms by exploiting the integrals (\ref{Ythere}). 
Then, we perform the integrations by parts with respect to $t$ and $r$ 
and use the background Eqs.~(\ref{back1})-(\ref{back3}) 
and (\ref{back4})-(\ref{back5}). After the lengthy but straightforward calculations, 
the reduced second-order action is expressed in the form 
\be
{\cal S}_{\rm even}^{(2)}=\sum_{l} \int \rd t \rd r {\cal L}_{\rm even}\,,
\label{Seven}
\ee
where
\ba
{\cal L}_{\rm even}
&=& 
H_0 \left[ a_1 \delta \phi'' +a_2 \delta \phi' +a_3 H_2'
+L a_4 h_1'+\left( a_5+L a_6 \right) \delta \phi 
+\left( a_7+L a_8 \right) H_2+L a_9 h_1 +a_{10} \delta \rho \right] 
\notag \\
& &
+(L b_1+\tilde{b}_1) H_1^2+H_1 \left( b_2 \dot{\delta \phi}'+b_3 \dot{\delta \phi}
+b_4 \dot{H}_2+L b_5 \dot{h}_1 \right)
+c_1 \dot{\delta \phi} \dot{H}_2
+H_2 \left[ c_2 \delta \phi'+ (c_3+L c_4) \delta \phi
+L c_5 h_1 \right]+c_6 H_2^2 
\notag \\
& &
+L d_1 \dot{h}_1^2+L h_1 \left( d_2 \delta \phi'
+d_3 \delta \phi \right)+L d_4 h_1^2
+e_1 \dot{\delta \phi}^2+e_2 \delta \phi'^2
+\left( e_3+L e_4 \right) \delta \phi^2
\notag\\
&&
+L f_1 v^2+f_2 \delta \rho^2+f_3\delta\ell'H_1+\delta\ell
(f_4\dot{\delta\rho}+f_5\dot{H}_2+f_6\delta u_r'+f_7\delta u_r+Lf_8v)
+f_9\delta u_r^2+f_{10}\delta u_r\delta{\cal A}_1+Lf_{11}v\delta{\cal A}_2\notag\\
&&
+f_{12}\delta{\cal A}_1H_1+f_{13}(\delta{\cal A}_1\delta\dot{\cal B}_1
+L\delta{\cal A}_2\delta\dot{\cal B}_2)
\notag\\
&&
+g_1  ( L\dot{G}'-2 \dot{K}' ) H_1+g_2  ( LG''-2 K'' ) H_0+g_3  ( L{\dot{G}}^{2}+2 {\dot{K}}^{2} ) +g_4  ( L{G'}^{2}+2 {K'}^{2} ) +Lg_5 \dot{G} \dot{K}+Lg_6 G' K'
\notag\\
&&
+ ( L\dot{G}-2 \dot{K} )  ( g_7 \dot{H}_2+g_8 \dot{\delta\phi}+g_9 H_1+g_{10} \delta\ell ) 
+Lg_{11}  ( \dot{G}-\dot{K} ) h_0
\notag\\
&&
+ ( LG'-2 K' )  ( g_{12} \delta\phi'+g_{13} H_0+g_{14} H_2+g_{15} \delta\phi ) 
+Lg_{16}  ( G'-K' ) h_1+ ( L-2 ) K ( k_1 H_0+k_2 H_2+k_3 \delta\phi ) \notag\\
&&
+L [ m_1 h_0'^{2}+m_2 h_0'  ( \dot{h}_1+H_1 ) +m_3 h_0^{2}+h_0  \{ m_4 \dot{H}_2+m_5 \dot{h}_1+m_6 \dot{\delta\phi}+m_7 H_1+m_8  ( \delta\ell+\delta {\cal A}_2  )  \}  ] 
\,.
\label{eaction}
\ea
The coefficients $a_1, a_2, ...$ etc, which depend on the background quantities, 
are explicitly given in Appendix A\footnote{By using the Maple software, we confirmed 
that the calculations independently performed by the authors agree with each other.}.

\subsection{Perturbation equations of motion}
\label{linearpersec}

{}From the action (\ref{Seven}), we will derive the full linear perturbation 
equations of motion without fixing any particular gauge conditions.
For the coefficients 
$a_{10}$, $\tilde{b}_1$, $f_i$ $(i=1,...,13)$, $g_{10}$, and $m_8$,
we use their explicit forms given in Appendix A.

We first derive the perturbation equations associated with the 
perfect fluid. Varying the action (\ref{Seven})
with respect to $v$, we obtain
\be
\delta\ell=\frac{\rho+P}{n}v-\delta {\cal A}_2\,.
\label{soldl}
\ee
This relation will be used to eliminate $\delta\ell$ and its derivatives from 
the other perturbation equations. 
Variation of the action (\ref{Seven}) with respect to $\delta\ell$, $\delta\rho$, $\delta u_r$, 
$\delta {\cal A}_1$, $\delta {\cal A}_2$, $\delta {\cal B}_1$, and $\delta {\cal B}_2$ 
lead, respectively, to
\ba
&&
{\frac {\dot{\delta\rho}}{\rho+P}}
+\frac12\dot{H}_2
+\dot{K}
-\frac12 L\dot{G}
-hH_1'
+h\sqrt {f}\,\delta u_r'
+ \left( \frac12 {\frac {f' h}{fc_m^2}}-\frac{h'}{2}-{\frac {2h}{r}} \right) H_1
+{\frac {L}{r^2}}h_0\notag\\
&&
+ \left[ \frac12 {\frac {h ( c_m^2-1 ) f'}{\sqrt {f}c_m^2}}
+\frac{\sqrt {f}h'}{2}+ {\frac {2h\sqrt {f}}{r}} \right] \delta u_r
-{\frac {L\sqrt {f}}{r^2}}v=0\,,\label{eqdl}\\
&&
{\frac {c_m^2 }{\rho+P}}\delta\rho
-\frac12H_0+{\frac {\dot{v}}{\sqrt {f}}}-{\frac {n\delta\dot{\cal A}_2}{\sqrt {f} ( \rho+P ) }}=0\,,\label{eqdrho}\\
&&\delta u_r-v'+{\frac {f' }{2f}}v-{\frac {n}{\rho+P}}( \delta{\cal A}_1-\delta{\cal A}_2' )=0\,,\label{eqdur}\\
&&\delta\dot{\cal B}_1+h ( \sqrt {f}\delta u_r-H_1 ) =0\,,\label{eqA1}\\
&&\delta\dot{\cal B}_2-{\frac {h_0-v\sqrt {f}}{r^2}}=0\,,\label{eqA2}\\
&&\delta\dot{\cal A}_1=0\,,\label{eqB1}\\
&&\delta\dot{\cal A}_2=0\,.\label{eqB2}
\ea
{}From Eqs.~(\ref{eqB1}) and (\ref{eqB2}), the Lagrange multipliers 
$\delta {\cal A}_1$ and $\delta {\cal A}_2$ depend on the distance $r$ alone. 
The last term on the left hand-side of Eq.~(\ref{eqdrho}) vanishes 
due to the property $\delta\dot{\cal A}_2=0$. 
In the following, we also employ the relations (\ref{eqB1}) and (\ref{eqB2}) 
for other perturbation equations of motion.
Taking the time derivative of Eq.~(\ref{eqdur}) and differentiating 
the resulting equation with respect to $r$ further, we have 
\ba
& &
\dot{\delta u}_r-\dot{v}'+\frac{f'}{2f} \dot{v}=0\,,
\label{delueq1}\\
& &
\dot{\delta u}_r'-\dot{v}''+\frac{f'}{2f} \dot{v}'
+\frac12\left(\frac{f''}{f}-\frac{f'^2}{f^2}\right)\dot{v}=0\,.
\label{delueq2}
\ea
The Lagrange multipliers $\delta {\cal B}_1$ and $\delta {\cal B}_2$ are 
related to the components of fluid four velocity  
through Eqs.~(\ref{eqA1}) and (\ref{eqA2}).

Varying the action (\ref{Seven}) with respect to the metric perturbations 
$H_0$, $H_1$, $H_2$, $h_0$, $h_1$ and $K$, the resulting perturbation 
equations are given, respectively, by 
\ba
&&
a_1 \delta\phi''+a_2 \delta\phi'+a_3 H_2'+La_4 h_1'
 + ( a_5+La_6 ) \delta\phi+ ( a_7+La_8 ) H_2+La_9 h_1
 +\frac12 {\frac {r^2\sqrt {f}}{\sqrt {h}}}\delta\rho\notag\\
&&
+g_2(L G''-2 K'')+g_{13}(LG'-2 K')+ (  L-2 )k_1 K=0\,,
\label{eqH0}\\
&&
\left[ 2 Lb_1-{\frac {r^2 ( \rho+P ) \sqrt {h}}{\sqrt {f}}} \right] H_1
+b_2 \dot{\delta\phi}'+b_3 \dot{\delta\phi}+b_4 \dot{H}_2
+Lb_5 \dot{h}_1 
+r^2 ( \rho+P ) \sqrt {h}\, \delta u_r
\notag\\
&&
+g_1(L \dot{G}'-2 \dot{K}')+g_9(L\dot{G} -2 \dot{K})+Lm_2 h_0'+Lm_7h_0 =0\,,
\label{eqH1}\\
&&-c_1 \ddot{\delta\phi} -b_4\dot{H}_1 
+c_2 \delta\phi' + ( c_3+Lc_4 ) \delta\phi+Lc_5 h_1
+2 c_6 H_2-a_3H_0'+ ( La_8+a_7-a_3' ) H_0
-\frac12 {\frac {r^2 ( \rho+P ) }{\sqrt {h}}}\dot{v}
\notag\\
&&
-g_7(L\ddot{G}-2 \ddot{K})+g_{14}(LG' -2 K' )+ k_2(  L-2  ) K-Lm_4\dot{h}_0 =0\,,
\label{eqH2}\\
&&-2 Lm_1 h_0''-2 Lm_1' h_0'-Lm_2 (\dot{h}_1'+H_1')+2 Lm_3h_0 +Lm_4\dot{H}_2
+ ( m_5-m_2' ) L\dot{h}_1+Lm_6 \dot{\delta\phi}+ ( m_7-m_2' ) LH_1
\notag\\
&&
+{\frac { ( \rho+P ) L}{\sqrt {h}}}v+Lg_{11}( \dot{G}-\dot{K})=0\,,
\label{eqh0}\\
&&-2 d_1 \ddot{h}_1+d_2 \delta\phi'+d_3 \delta\phi+2 d_4 h_1
-a_4H_0' +( a_9-a_4' ) H_0-b_5\dot{H}_1+ c_5 H_2
-m_2 \dot{h}_0' -m_5 \dot{h}_0+g_{16}(G'-K')=0\,,\qquad
\label{eqh1}\\
&&
-2 g_1 \dot{H}_1'-2 g_2 H_0''-4 g_3 \ddot{K}-4 g_4 K''-4 g_4' K'-Lg_5 \ddot{G}
-Lg_6 G''-Lg_6' G'+2 g_7 \ddot{H}_2+2 g_8 \ddot{\delta\phi}+ 2( g_9- g_1' ) \dot{H}_1
\notag\\
&&
+Lg_{11} \dot{h}_0+2 g_{12} \delta\phi''
+ 2(  g_{13}-2 g_2' ) H_0'+2 g_{14} H_2'+ 2(  g_{15}+ g_{12}' ) \delta\phi'
+Lg_{16} h_1'+Lg_{16}' h_1
\notag\\
&&
+ [ (L-2)k_3 +2 g_{15}' ] \delta\phi+ [ (L-2)k_1 -2 g_2''+2 g_{13}' ] H_0
+ [ (L-2)k_2 +2 g_{14}' ] H_2-{\frac {r^2 ( \rho+P )}{\sqrt {h}}}\dot{v}=0\,.
\label{eqK}
\ea
For the derivation of Eq.~(\ref{eqH1}), we used Eq.~(\ref{eqdur}) 
to eliminate the combination $\delta{\cal A}_1-\delta{\cal A}_2'$ 
of Lagrange multipliers.

Varying the action (\ref{Seven}) with respect to $G$ and combining the 
resulting equation with Eq.~(\ref{eqK}), it follows that 
\ba
 &&
 -2( 2 g_3+ g_5 ) \ddot{K} - ( 4 g_3+Lg_5 ) \ddot{G}
 -2 ( 2 g_4+ g_6 ) K''- ( 4 g_4+Lg_6 ) G''
 -2 ( 2 g_4'+ g_6' ) K'- ( 4 g_4'+Lg_6' ) G'
 \notag\\
 &&
 +( L-2 )\left( g_{11} \dot{h}_0+g_{16} h_1'+g_{16}'  h_1
+k_1 H_0+k_2 H_2+k_3 \delta\phi \right)=0\,.
\label{eqG}
\ea

Finally, variation of the action (\ref{Seven}) with respect to $\delta\phi$ 
leads to 
\ba
&&
-2 e_1 \ddot{\delta\phi}-2 e_2 \delta\phi''+ 2( e_3+Le_4 ) \delta\phi
+a_1H_0'' + ( 2 a_1'-a_2 ) H_0'+ ( a_1''-a_2'+a_5+La_6 ) H_0
+b_2\dot{H}_1' + ( b_2'-b_3 ) \dot{H}_1
\notag\\
&&
-c_1 \ddot{H}_2-c_2H_2' - ( c_2'-c_3-Lc_4 ) H_2
-Ld_2h_1' +L ( d_3-d_2' ) h_1-2 e_2' \delta\phi'
+g_8(2 \ddot{K} -L\ddot{G})+g_{12}(2 K'' -LG'' )
\notag\\
&&
-( g_{15}- g_{12}' ) (2K'-L G')+ k_3(  L-2 ) K-Lm_6\dot{h}_0 =0\,.
\label{eqdphi}
\ea
We thus derived the full set of linear perturbation equations of motion 
in the gauge-ready form. 
We note that Eqs.~(\ref{eqH0})-(\ref{eqdphi}) do not contain 
the Lagrange multipliers $\delta {\cal A}_i$ and $\delta {\cal B}_i$. 

%%%%%%%%%%%%%%%%%%%%%%%%%%%%%%%%%%%%%%%%%%%
\section{Propagation of even-parity perturbations}
\label{cssec}
%%%%%%%%%%%%%%%%%%%%%%%%%%%%%%%%%%%%%%%%%%%

To study the propagation of even-parity perturbations, we fix the residual 
gauge degrees of freedom. We consider  
the infinitesimal gauge transformation
$x_\mu\to x_\mu+\xi_\mu$, where 
\be
\xi_t=\sum_{l,m} {\cal T}(t,r)Y_{lm}(\theta,\varphi)\,,\qquad 
\xi_r=\sum_{l,m} {\cal R}(t,r)Y_{lm}(\theta,\varphi)\,,\qquad 
\xi_a=\sum_{l,m} \Theta(t,r) \nabla_a Y_{lm}(\theta,\varphi)\,.
\label{xivec2}
\ee
Then, the transformations of even-parity perturbations 
$H_0$, $H_1$, $H_2$, $h_0$, $h_1$, $K$, $G$, and 
$\delta \phi$ are given, respectively, by 
\cite{Kobayashi:2014wsa,Motohashi:2011pw}
\ba
& &
H_0 \to H_0+\frac{2}{f} \dot{{\cal T}}-\frac{f' h}{f}{\cal R}\,,\qquad 
H_1 \to H_1+\dot{{\cal R}}+{\cal T}'-\frac{f'}{f}{\cal T}\,,\qquad
H_2 \to H_2+2h{\cal R}'+h' {\cal R}\,,\label{H1tra} \nonumber \\
& &
h_0 \to h_0+{\cal T}+\dot{\Theta}\,,\qquad 
h_1 \to h_1+{\cal R}+\Theta'
-\frac{2}{r}\Theta\,,\qquad
K \to K+\frac{2}{r}h{\cal R}\,,\qquad 
G \to G+\frac{2}{r^2}\Theta\,,\label{beal} \nonumber \\
& &
\delta \phi \to \delta \phi-\phi' h{\cal R}\,.\label{gauge}
\ea
It is possible to construct gauge-invariant quantities invariant under the 
above gauge transformation. 
One can combine $G$ and its derivatives with other perturbed quantities 
to eliminate $\Theta$. 
In the same manner, the transformation scalar ${\cal T}$ 
can be removed by combining perturbed quantities with 
$h_0-r^2\dot{G}/2$ and its derivatives. 
For the remaining transformation scalar ${\cal R}$, 
there are several ways to eliminate it. 
If we use $K$, for instance, we can construct the several  
gauge-invariant variables:
\ba
&&
\tilde{H}_0=
H_0
-\frac{2}{f}\frac{\partial}{\partial t}\left(h_0-\frac{r^2}{2}\dot{G}\right)
+\frac{rf'}{2f}K
\,,\qquad 
\tilde{H}_1=
H_1
-f\left[\frac{1}{f}\left(h_0-\frac{r^2}{2}\dot{G}\right)\right]'
-\frac{r}{2h}\dot{K}\,,
\notag\\
&&
\tilde{H}_2=
H_2
-\sqrt{h}\left(\frac{rK}{\sqrt{h}}\right)'\,,\qquad
\tilde{h}_1=
h_1
-\frac{r^2}{2}G'
-\frac{r}{2h}K\,,\qquad
\tilde{\delta \phi}=
\delta \phi
+\frac{r\phi'}{2}K\,.
\label{invariants}
\ea
In the later discussion, we will consider the following 
gauge-invariant combination
\be
\psi = \tilde{H}_2+\frac{a_4}{a_3}L\tilde{h}_1+\frac{a_1}{a_3}\tilde{\delta\phi}'\,.
\label{psig}
\ee

For $l \geq 2$, we can fix the transformation scalars 
${\cal T}$ and $\Theta$ by choosing the gauge:
\be
h_0=0\,,\qquad G=0\,.
\label{gauge1}
\ee
In order to fix the other transformation scalar ${\cal R}$, 
Regge and Wheeler \cite{Regge:1957td} chose the spatially diagonal gauge 
$h_1=0$. The other possible choices 
are the uniform curvature gauge $K=0$
and the unitary gauge $\delta \phi=0$. 
Following the references \cite{DeFelice:2011ka,Motohashi:2011pw,Kobayashi:2014wsa} 
of even-parity perturbations in modified gravity theories,
we choose the gauge 
\be
K=0\,.
\label{gauge2}
\ee
Then, the gauge-invariant variables in Eq.~(\ref{invariants}) reduce to 
$\tilde{H}_2=H_2$, $\tilde{h}_1=h_1$, and $\tilde{\delta\phi}=\delta\phi$. 
For this gauge choice, the quantity (\ref{psig}) yields
\be
\psi = H_2+\frac{a_4}{a_3}Lh_1+\frac{a_1}{a_3}\delta\phi'\,,
\label{psidef}
\ee
which corresponds to that introduced in Ref.~\cite{Kobayashi:2014wsa}.

In Horndeski theories with matter, 
there are three propagating degrees of freedom in the even-parity sector. 
The first is the fluid density perturbation $\delta\rho$, and 
the second is the scalar-field perturbation $\delta\phi$. 
The third is the perturbation $\psi$, which corresponds to 
the propagating degree of freedom arising from the gravity sector.
{}From Eq.~(\ref{psidef}), we replace $H_2$ with $\psi$ to study the 
propagation of the dynamical degrees of freedom.
We derive second-order differential equations of 
the three dynamical variables $\delta\rho$, $\delta\phi$, $\psi$ 
and study the propagation of them along the radial and 
angular directions.  
For this purpose, we need to eliminate the other non-dynamical variables 
from the perturbation equations presented in Sec.~\ref{linearpersec}. 
In the following, we first investigate the case $l \geq 2$ and then finally 
study the $l=0$ and $l=1$ cases separately.

\subsection{$l \geq 2$}

To reduce the number of nondynamical degrees of freedom from 
the perturbation equations of motion for $l \geq 2$, 
we first take the time derivative of Eq.~(\ref{eqdl}) and eliminate 
$\dot{\delta u}_r$ and $\dot{\delta u}_r'$ on account of 
Eqs.~(\ref{delueq1}) and (\ref{delueq2}). 
This process gives rise to the third derivative $\dot{v}''$ 
as well as $\dot{v}'$, $\dot{v}$. 
{}From Eqs.~(\ref{eqdrho}) and (\ref{eqB2}), 
we have 
\be
\dot{v}=-\sqrt{f} \left( \frac{c_m^2}{\rho+P} 
\delta \rho-\frac12 H_0 \right)\,.
\label{dotveq}
\ee
Taking the $r$ derivative of this relation twice, 
the third derivative $\dot{v}''$ is replaced with the 
second derivatives $\delta \rho''$ and $H_0''$.
Indeed, this procedure generates a Laplacian term 
proportional to $c_m^2 \delta \rho''$. 
As a consequence, we obtain the perturbation equation 
containing up to the second derivatives of $\delta \rho$, as
\ba
\hspace{-1.1cm}
&&
{\frac {2\ddot{\delta\rho}}{\rho+P}}+\ddot{\psi}-{\frac {2fhc_m^2}{\rho+P}}\delta\rho''
- \left[ 8 fhc_mc_m'+ ( 4 c_m^2+1 ) hf'+c_m^2f h'+{\frac {4fhc_m^2}{r}} \right] 
{\frac {\delta\rho'}{\rho+P}}
\notag\\
\hspace{-1.1cm}
&&
+ \left[ {\frac {2fc_m^2 L}{r^2}}-4 fh(c_mc_m''+c_m'^2)
- 2fh\left( {\frac {4f'}{f}}+{\frac {h'}{h}}+\frac{4}{r} \right) c_mc_m'
-\frac12 f' h   \left( {\frac {f'}{f}}+{\frac {h'}{h}}+\frac{4}{r}
+{\frac {2f''}{f'}} \right)( 1+c_m^2 )  \right] }{\frac {\delta\rho}{\rho+P}
\notag\\
\hspace{-1.1cm}
&&
-{\frac {La_4 \ddot{h}_1}{a_3}}
+ \left( {\frac {f' h}{fc_m^2}}-h'- {\frac {4h}{r}} \right) \dot{H}_1
+ \left[  {\frac {h ( 2 c_m^2-1 ) f'}{2c_m^2}}+\frac{fh'}{2} +{\frac {2fh}{r}} \right] H_0'
-{\frac {f}{r^2}}LH_0-{\frac {a_1 \ddot{\delta\phi}'-ha_3  ( fH_0''-2 \dot{H}_1' ) }{a_3}}=0,
\label{eqdlu}
\ea
where we used the property $\rho'=(\rho+P)n'/n$ with $n'=-nf'/(2f c_m^2)$.
This equation still contains the third derivative $\ddot{\delta\phi}'$, 
but it can be removed by using other perturbation equations of motion. 
Indeed, the same higher-order term also appears in Eqs.~(\ref{eqK}) and (\ref{eqdphi}) 
from the term $\ddot{H}_2$ by taking the second time derivative of Eq.~(\ref{psidef}).
These equations are given, respectively, by 
\ba
&&
g_7 \ddot{\psi}+g_8 \ddot{\delta\phi}
-{\frac { a_1 g_{14}-a_3 g_{12} }{a_3}}\delta\phi''+g_{14} \psi'
- \left[ 
\frac12 {\frac {a_1 k_2 L}{a_3}}-g_{12}'-g_{15}
-{\frac { ( k_2-g_{14}' ) a_1-g_{14} a_1'}{a_3}}
-{\frac {a_1 a_3' g_{14}}{{a_3}^{2}}} \right] \delta\phi'
\notag\\
&&
+ \left( \frac12 k_2 L-k_2+g_{14}' \right) \psi
+ \left( \frac12 k_3 L-k_3+g_{15}' \right) \delta\phi
+{\frac {c_m^2 r^2\sqrt {f}}{2\sqrt {h}}}\delta\rho
-{\frac {g_7 a_4}{a_3}}L\ddot{h}_1
+ ( g_9-g_1' ) \dot{H}_1+ ( g_{13}-2 g_2' ) H_0'
\notag\\
&&
- \left( {\frac {g_{14} a_4}{a_3}}-\frac{g_{16}}{2} \right) Lh_1'
+ \left[ \frac12 k_1 L-k_1+g_{13}'-g_2''-{\frac { ( \rho+P ) r^2\sqrt {f}}{4\sqrt {h}}} \right] H_0
\notag\\
&&
- \left[ \frac12 {\frac {a_4 k_2 L}{a_3}}
-\frac{g_{16}'}{2}+{\frac {( g_{14}'-k_2 ) a_4+g_{14} a_4'}{a_3}}-{\frac {g_{14} a_3' a_4}{{a_3}^{2}}} \right] Lh_1
-\frac12 {\frac {r^2a_4  [ a_1 \ddot{\delta\phi}'-ha_3  ( fH_0''-2 \dot{H}_1' ) ] }{fha_3}}=0\,,
\label{eqpsiu}\\
&&
e_1 \ddot{\delta\phi}+\frac{c_1}{2} \ddot{\psi}+ 
\left( e_2-\frac12 {\frac {c_2 a_1}{a_3}} \right) \delta\phi''
+ \left[ \frac12 {\frac {c_4 a_1 L}{a_3}}+e_2'
+\frac12 {\frac { ( c_3-c_2' ) a_1-c_2 a_1'}{a_3}}+\frac12 {\frac {c_2 a_1 a_3'}{{a_3}^{2}}} \right] \delta\phi'
\notag\\
&&
+\frac{c_2}{2} \psi'
- ( e_3+Le_4 ) \delta\phi
-\frac12  ( Lc_4+c_3-c_2' ) \psi
-{\frac {c_1 a_4 }{2a_3}}L\ddot{h}_1
+\frac12  ( b_3-b_2' ) \dot{H}_1
+ \left( \frac{a_2}{2}-a_1' \right) H_0'
\notag\\
&&
+\frac12  \left( d_2-{\frac {c_2 a_4}{a_3}} \right) Lh_1'
-\frac12  ( La_6+a_5-a_2'+a_1'' ) H_0
\notag\\
&&
+\frac12  \left[ {\frac {c_4 a_4 L}{a_3}}-d_3
+d_2'+{\frac { ( c_3-c_2' ) a_4-c_2 a_4'}{a_3}}+{\frac {c_2 a_3' a_4}{{a_3}^{2}}} \right] Lh_1
+\frac12 {\frac {a_1  [ a_1 \ddot{\delta\phi}'-ha_3  ( fH_0''-2 \dot{H}_1' ) ] }{fha_3}}=0\,,
\label{eqdphiu}
\ea
where we eliminated the term $\dot{v}$ in Eq.~(\ref{eqK}) 
by using Eq.~(\ref{dotveq}). 
Under the gauge choice (\ref{gauge1})-(\ref{gauge2}), 
Eqs.~(\ref{eqH0}), (\ref{eqH2}), (\ref{eqh1}), (\ref{eqG}) 
reduce, respectively, to 
\ba
&&
\left[ {\frac {a_4 a_8 L}{a_3}}-a_9+a_4'+{\frac {a_4  ( a_7-a_3' ) }{a_3}} \right] Lh_1
\notag\\
&&
=
\psi' a_3- \left[ {\frac {a_1 a_8 L}{a_3}}-a_2+a_1'+{\frac {a_1  ( a_7-a_3' ) }{a_3}} \right] 
\delta\phi'+ ( La_6+a_5 ) \delta\phi+ ( La_8+a_7 ) \psi+\frac12 {\frac {r^2\sqrt {f}\delta\rho}{\sqrt {h}}} 
\,,\label{eqH0u}\\
&&
{\frac {a_1 \ddot{\delta\phi}-ha_3  ( fH_0'-2 \dot{H}_1 ) }{fh}}
+ \left[ La_8-a_3'+a_7-{\frac { ( \rho+P ) r^2\sqrt {f}}{4\sqrt {h}}} \right] H_0
+ \left( c_5-{\frac{2c_6 a_4}{a_3}} \right) Lh_1
\notag\\
&&
=
\left( {\frac {2c_6 a_1}{a_3}}-c_2 \right) \delta\phi'-( Lc_4+c_3 ) \delta\phi-2 c_6 \psi
- {\frac {c_m^2 r^2\sqrt {f}\delta\rho}{2\sqrt {h}}}
\,,\label{eqH2u}\\
&&
{\frac {a_4}{f}}(\ddot{h}_1+fH_0'-\dot{H}_1)+ ( a_4'-a_9 ) H_0
+\left( {\frac {a_4c_5 L}{a_3}}-2 d_4 \right) h_1
=\left( d_2-{\frac {a_1 c_5}{a_3}} \right) \delta\phi'
+d_3 \delta\phi+\psi c_5\,,
\label{eqh1u}\\
&&
k_1 H_0+g_{16} h_1'-\left({\frac {a_4 k_2 L}{a_3}}-g_{16}' \right) h_1
=
{\frac {a_1 k_2 }{a_3}}\delta\phi'-k_3 \delta\phi-k_2\psi\,.
\label{eqGu}
\ea

Now, we will derive the second-order coupled differential equations 
for $\delta \rho$, $\psi$, and $\delta \phi$ by eliminating 
nondynamical variables $H_0$, $H_1$, $H_2$, $h_1$ 
and their derivatives present in Eqs.~(\ref{eqdlu})-(\ref{eqGu}).
The second time derivative of $h_1$ in Eqs.~(\ref{eqdlu})-(\ref{eqdphiu}) 
can be eliminated by solving Eq.~(\ref{eqh1u}) for $\ddot{h}_1$. 
Taking the $r$ derivative of Eq.~(\ref{eqH2u}), we can  
remove the combination 
$a_1 \ddot{\delta\phi}'-ha_3 ( fH_0''-2 \dot{H}_1')$ appearing in Eqs.~(\ref{eqdlu})-(\ref{eqdphiu}).
After this procedure, we eliminate the term $fH_0'-2\dot{H}_1$ 
by using Eq.~(\ref{eqH2u}). 
The next step is to solve Eq.~(\ref{eqH0u}) for $h_1$ and to 
take its $r$ derivative. After substituting $h_1$ and $h_1'$ into 
Eq.~(\ref{eqGu}), we solve this equation for $H_0$.
On using these relations, we can remove the nondynamical variables 
$h_1'$, $H_0$, and $h_1$ from Eqs.~(\ref{eqdlu})-(\ref{eqdphiu}).
Then, we finally end up with the second-order differential equations 
containing only the dynamical perturbations
$\delta\rho$, $\psi$, $\delta\phi$ and their derivatives. 
This is expressed in the form
\be
{\bm K}\ddot{\vec{\mathcal{X}}}
+{\bm G}\vec{\mathcal{X}}^{''}
+{\bm Q}\vec{\mathcal{X}}^{'}
+{\bm M} \vec{\mathcal{X}}=0\,,
\label{Xpereq}
\ee
where ${\bm K}$, ${\bm G}$, ${\bm Q}$, ${\bm M}$ 
are $3 \times 3$ matrices, with 
\be
\vec{\mathcal{X}}=\left( 
\begin{array}{c}
\delta \rho\\
\psi\\
\delta \phi 
\end{array}
\right) \,.
\label{calX}
\ee
In the following, we will investigate the propagation of dynamical 
perturbations in both radial and angular directions.

\subsubsection{Radial propagation}

The dispersion relation along the radial direction can be 
derived by assuming the solutions of Eq.~(\ref{calX}) to be
$\vec{\mathcal{X}}=\vec{\mathcal{X}}_0 e^{i( \omega t-kr)}$, 
where $\vec{\mathcal{X}}_0$ is a constant vector, and 
$\omega$ and $k$ are the constant frequency and wavenumber 
respectively. In the limits $\omega \to \infty$ and $k \to \infty$, 
the dominant contributions to Eq.~(\ref{Xpereq}) are the first 
two terms. To have nonvanishing solutions of 
$\vec{\mathcal{X}}$, we require that
${\rm det}|\omega^2{\bm K}+k^2 {\bm G}|=0$.
The radial propagation speed in the coordinates $r$ and $t$ 
is given by $\hat{c}_r={\rm d}r/{\rm d}t=\omega/k$.
We consider the propagation speed $c_r$ measured 
by the rescaled radial coordinate $r_*=\int {\rm d}r/\sqrt{h}$
and proper time $\tau=\int \sqrt{f}\,{\rm d}t$. 
In this case we have $c_r={\rm d}r_*/{\rm d}\tau=\hat{c}_r/\sqrt{fh}$,
so the dispersion relation translates to 
\be
{\rm det}|fhc_r^2{\bm K}+{\bm G}|=0\,.
\label{dispr}
\ee
The matrix components of ${\bm K}$ and $\bm{G}$ have the following 
properties
\ba
& &
K_{21}=0\,,\qquad K_{31}=0\,,
\label{K210}\\
& &
G_{21}=0\,,\qquad G_{31}=0\,.
\label{K21}
\ea
Then, the matter propagation speed $c_{r1}$ along the radial 
direction decouples from the other two, so that  
\be
fh c_{r1}^2K_{11}+G_{11}=0\,,
\ee
where
\be
K_{11}=\frac{2}{\rho+P}\,,\qquad 
G_{11}=-\frac{2fh}{\rho+P}c_m^2\,. 
\label{K11}
\ee
Hence we obtain
\be
c_{r1}^2=c_m^2\,.
\label{cr1m}
\ee
The matter perturbation $\delta \rho$ propagates with the 
sound speed $c_m$. This is different from the value 
$c_{r1}^2=0$ obtained in Ref.~\cite{Kase:2020qvz}, but 
the latter arises from inappropriate treatment for deriving a reduced 
matter Lagrangian. 
The correct radial propagation speed squared 
along the radial direction is given by Eq.~(\ref{cr1m}). 
In Appendix B, we will see that this difference 
stems from how to integrate out nondynamical perturbations 
from the action (analogous to the discussion of odd-parity perturbations
in Sec.~\ref{peroddsec}).

From Eq.~(\ref{dispr}), the other two propagation speeds 
are obtained by solving 
\be
(K_{22}K_{33}-K_{23}K_{32})f^2h^2c_r^4
+(K_{22}G_{33}+K_{33}G_{22}-K_{23}G_{32}-K_{32}G_{23})fhc_r^2
+G_{22}G_{33}-G_{23}G_{32}=0\,.
\label{detcr}
\ee
Although the nonvanishing components of matrices $\bm{K}$ and $\bm{G}$ 
are complicated, the combination of terms appearing in Eq.~(\ref{detcr}) 
can be simplified in the following way. 
On using the relations among the quantities given in Appendix~A, all of them can be 
expressed in terms of $a_1$, $a_4$, $c_2$, $c_3$, $c_4$, $e_3$, ${\cal F}$, ${\cal G}$, 
and the derivatives of them. In addition to these variables, 
we introduce the following combination \cite{Kobayashi:2014wsa,Kase:2020qvz}, 
\be
{\cal P}_1 \equiv \frac{h \mu}{2fr^2 {\cal H}^2} \left( 
\frac{fr^4 {\cal H}^4}{\mu^2 h} \right)'\,,
\label{defP1}
\ee
with 
\be
\mu=-\frac{4a_3}{\sqrt{fh}}
=\frac{2(\phi' a_1+2ra_4)}{\sqrt{fh}}\,,\qquad
{\cal H} = \frac{2a_4}{\sqrt{fh}}\,,
\ee
where ${\cal H}$ is the same quantity as defined in Eq.~(\ref{cHdef}). 
We will replace the derivative $a_1'$ with ${\cal P}_1$. 
{}From the background Eqs.~(\ref{back1}) 
and (\ref{back3}), there is the following relation
\be
a_4'={\frac {1}{2 f-rf'} \left[  \left( rf''-{\frac {r{f'}^{2}}{f}}+2 f'-{\frac {2f}{r}} \right) a_4
+{\frac {r{f}^{3/2}}{\sqrt {h}} \left( {\frac {{\cal F}}{r^2}}-\rho-P \right) } \right] }\,, 
\label{conda4}
\ee
which will be used to eliminate the term $a_4'$.
As a result, Eq.~(\ref{detcr}) is factorized in the form, 
\be
(c_r^2-c_{r2}^2)(c_r^2-c_{r3}^2)=0\,,
\ee
whose solutions to $c_r^2$ are given by 
\ba
c_{r2}^2 &=&
\frac{\cal G}{\cal F}\,,
\label{cr0}\\
c_{r3}^2 &=&
{\frac { 4\phi'[ 8r^2 f h a_4 c_4 (\phi' a_1+ra_4)
-a_1^{2}{f}^{3/2}\phi' {\cal G} \sqrt {h}+2  ( a_1 f'+2 c_2 f ) a_4^{2}r^2] }
{f^{5/2}\sqrt {h}[(2{\cal P}_1-{\cal F})
h \mu^2 -2{\cal H} ^2 r^4 (\rho+P) ]}}
\,.\label{cr}
\ea
The value $c_{r2}$, which is equivalent to the radial propagation speed 
(\ref{crodd}) in the odd-parity sector, corresponds to the 
speed of gravitational perturbation $\psi$. 
We also note that $c_{r2}$ coincides with the one derived 
in Ref.~\cite{Kobayashi:2014wsa}.
The other value $c_{r3}$ is the propagation speed 
of scalar-field perturbation $\delta \phi$. 
The presence of matter affects $c_{r3}$ through the 
term proportional to $\rho+P$ in the denominator of Eq.~(\ref{cr}). 
The radial Laplacian stabilities of $\psi$ and $\delta \phi$ 
are ensured under the two conditions $c_{r2}^2 \geq 0$ and $c_{r3}^2 \geq 0$. 
We note that the radial speed squares (\ref{cr0}) and (\ref{cr}) are 
different from those derived in Ref.~\cite{Kase:2020qvz}. 
In Ref.~\cite{Kase:2020qvz}, the propagations of $\psi$ and 
$\delta \phi$ are affected by inappropriate treatment for the 
integration of the Schutz-Sorkin action. 
Now, we corrected the values of $c_{r2}^2$ and $c_{r3}^2$ 
together with the matter propagation speed squared $c_{r1}^2$. 
 
Since the matter perturbation $\delta \rho$ is decoupled from $\psi$ 
and $\delta \phi$ for high frequency modes, 
the no-ghost condition for $\delta \rho$ corresponds to 
the standard weak energy condition, i.e., 
\be
\rho+P>0\,,
\label{rhoPcon}
\ee
under which $K_{11}$ is positive. 
The quantities associated with the no ghost conditions of 
other two dynamical perturbations are given by 
\ba
K_{22} &=& \frac{r^2{\cal H}}{4\sqrt{fh}}\,,\\ 
K_{22}K_{33}-K_{23}K_{32} &=&
\frac{r}{8 fh^2 \phi'^2 \mu} \left[ (2{\cal P}_1-{\cal F})
h \mu^2 -2{\cal H} ^2 r^4 (\rho+P) \right]\,.
\label{calK}
\ea
The stability of odd-parity perturbations requires that 
${\cal H}>0$, under which $K_{22}$ is positive.
Since there are freedoms to multiply any positive or negative
quantities to the third equation of (\ref{Xpereq}), the 
sign of $K_{22}K_{33}-K_{23}K_{32}$ does not necessarily fix the
no-ghost condition.
Without the perfect fluid, however, we know that the ghost is 
absent for $2{\cal P}_1-{\cal F}>0$ \cite{Kobayashi:2014wsa}, 
whose combination appears in Eq.~(\ref{calK}).
Provided that the terms inside the square bracket of 
Eq.~(\ref{calK}) are positive, i.e., 
\be
{\cal K} \equiv
(2{\cal P}_1-{\cal F})
h \mu^2 -2{\cal H} ^2 r^4 (\rho+P)>0\,,
\label{noghost}
\ee
we recover the no-ghost condition without matter.
Then the inequality (\ref{noghost}) can be regarded as 
the no-ghost condition in the presence of matter. 
Under the condition (\ref{noghost}) the denominator 
of Eq.~(\ref{cr}) is positive, so the inequality $c_{r3}^2 \geq 0$ requires 
that the numerator of Eq.~(\ref{cr}) is nonnegative.
In summary, the absence of ghosts in the even-parity sector 
imposes the two additional conditions (\ref{rhoPcon}) and (\ref{noghost}).

\subsubsection{Angular propagation}

Let us proceed to the derivation of propagation speeds along the 
angular direction. In doing so, we substitute the solution of the form 
$\vec{\mathcal{X}}=\vec{\mathcal{X}}_0
e^{i(\omega t-l \theta)}$ into the perturbation equations 
(\ref{Xpereq}). In the limits of large $\omega$ and $l$, 
the dispersion relation is given by ${\rm det}|\omega^2{\bm K}-{\bm M}|=0$. 
The angular propagation speed $\hat{c}_{\Omega}=r{\rm d}\theta/{\rm d}t$  
satisfies $\omega^2=\hat{c}_{\Omega}^2L/r^2$ in the limit $l \to \infty$. 
The speed $c_{\Omega}$ measured in proper time $\tau$ is 
$c_{\Omega}=r {\rm d}\theta/{\rm d}\tau=
\hat{c}_{\Omega}/\sqrt{f}$, so that 
$\omega^2=c_{\Omega}^2f L/r^2$.
Then, we can derive $c_{\Omega}^2$ by solving
\be
{\rm det}|fLc_{\Omega}^2{\bm K}-r^2{\bm M}|=0\,. 
\label{dispo}
\ee
Since the components $M_{21}$ and $M_{31}$ of matrix ${\bm M}$ 
do not vanish, the matter propagation speed does not apparently 
decouple from other speeds of propagation. 
However, we will see that the former eventually reduces to $c_m^2$ 
after the following manipulation. 
Expanding the matrix components of ${\bm M}$ in the limit $L \to \infty$, 
the leading-order components of $M_{1i}$, $M_{2i}$ (where $i=1,2,3$) 
have the dependence linearly in $L$, whereas $M_{3i}$'s have 
the leading-order terms proportional to $L^2$. 
Substituting the expanded components of ${\bm M}$ into Eq.~(\ref{dispo}), 
the leading-order terms are in proportion to $L^4$. 
However, they vanish identically after using the relations among the coefficients 
given in Appendix~A together with Eqs.~(\ref{defP1}) and (\ref{conda4}). 
Then, we need to pick up the next-to leading-order terms 
proportional to $L^3$.
Then, the dispersion relation (\ref{dispo}) gives 
the following equation
\be
(c_{\Omega}^2-c_m^2)(c_{\Omega}^4+2B_1c_{\Omega}^2+B_2)=0\,,
\label{cOm}
\ee
where 
\ba
\hspace{-0.8cm}
&&
B_1=
{\frac { a_4 \sqrt {h}r^3 [ 4 h ( \phi' a_1+2 ra_4 ) \beta_1+\beta_2
-4 \phi' a_1 \beta_3+2 fra_4 {\cal G}  ( \rho+P )  ] 
-2 fh^{3/2}{\cal G}  [ 2 ra_4  ( 2 {\cal P}_1-{\cal F} )  ( \phi' a_1+ra_4 ) 
+\phi'^2a_1^{2}{\cal P}_1 ] }
{ 4\sqrt {f} a_4[( \phi' a_1+2 ra_4 ) ^{2} 
( 2 {\cal P}_1-{\cal F} ) h-2 r^4a_4^{2} ( \rho+P )]}},
\label{B1def}
\hspace{-0.8cm}\nonumber \\
\hspace{-0.8cm}\\
\hspace{-0.8cm}
&&
B_2=
-r^2{\frac {r^2h \beta_1 [ 2 fh {\cal F} {\cal G} ( \phi' a_1+2 ra_4 ) +r^2\beta_2 ] 
-{r}^{4}\beta_2 \beta_3
-fh{\cal F} {\cal G}  ( \phi' fh {\cal F} {\cal G}a_1 +4 r^3 a_4 \beta_3 ) }
{ f{\cal F} \phi' a_1[  ( \phi' a_1+2 ra_4 ) ^{2} ( 2 {\cal P}_1-{\cal F} ) h-2 r^4a_4^{2} ( \rho+P )  ] }}\,,
\label{B2def}
\ea
with 
\ba
\beta_1&=&\phi'^2a_4 e_4-2 \phi' c_4 a_4'
+ \left[  \left( {\frac {f'}{f}}+{\frac {h'}{h}}-\frac{2}{r} \right) a_4
+{\frac {\sqrt {fh}{\cal F}}{r}} \right] \phi' c_4+{\frac {f{\cal F} {\cal G}}{2r^2}}\,,\\
\beta_2&=& \left[ {\frac {\sqrt {fh}{\cal F}}{r^2} 
\left( 2 hr\phi'^2c_4+{\frac {rf' \phi'a_4}{f}}-\sqrt {fh}\phi'{\cal G} \right) }
-{\frac {2\sqrt {fh}\phi' a_4 {\cal G}}{r} 
\left( {\frac {{\cal G}'}{{\cal G}}}-{\frac {a_4'}{a_4}}+{\frac {f'}{f}}+\frac12 {\frac {h'}{h}}-\frac{1}{r} \right) } \right] a_1-{\frac {4{\cal F} {\cal G} fha_4}{r}}\,,\qquad\,\,\\
\beta_3&=& \left( hc_4'-\frac{d_3}{2}+\frac12 h' c_4 \right) \phi' a_4
+ \left( {\frac {h'}{2h}}-\frac{1}{r}+{\frac {f'}{2f}}-{\frac {a_4'}{a_4}} \right)  
\left( {\frac {a_4 f'}{2f}}+2 h \phi'c_4+{\frac {\sqrt {fh}{\cal G}}{2r}} \right) a_4\notag\\
&&
+{\frac {\sqrt {fh}{\cal F}}{4r} \left( {\frac {f'}{f}}a_4+2 h \phi'c_4+{\frac {3\sqrt {fh}{\cal G}}{r}} \right) }\,.
\ea
In deriving these coefficients,  we replaced $f''$ with $a_4'$ on account of 
Eq.~(\ref{conda4}).

One of the solutions to Eq.~(\ref{cOm}) corresponds to the matter 
propagation speed squared, i.e., 
\be
c_{\Omega1}^2=c_m^2\,. 
\ee
Hence, as in the case of radial mode, the propagation speed of matter along 
the angular direction is decoupled from those of other perturbations.
The other two propagation speeds $c_{\Omega\pm}$ are given by 
\be
c_{\Omega\pm}^2=-B_1\pm\sqrt{B_1^2-B_2}\,.
\label{cosq}
\ee
Since there exist terms proportional to $\rho+P$ 
in the denominators of $B_1$ and $B_2$, the angular propagations of 
$\psi$ and $\delta \phi$ are generally affected by the presence of matter. 
The angular Laplacian instabilities of $\psi$ and $\delta \phi$ are absent 
under the two conditions $c_{\Omega+}^2 \geq 0$ and 
$c_{\Omega-}^2 \geq 0$.

\subsection{$l=0$}

For the monopole mode ($l=0$), the terms containing $h_0$, $h_1$, and $G$ 
in the second-order Lagrangian (\ref{eaction}) vanish identically. 
We choose the gauge $K=0$ to fix the residual gauge degree of freedom. 
We can exploit the perturbation 
Eqs.~(\ref{eqdlu})-(\ref{eqGu}) by setting $h_1$ and its 
derivatives 0, with $L=0$. 
To eliminate nondynamical perturbations, we first solve 
Eqs.~(\ref{eqH2u}) and (\ref{eqGu}) for $H_0'$ and 
$H_0$ respectively.
We take the $r$ derivative of Eq.~(\ref{eqH2u}) and solve 
it for $\ddot{\delta \phi}'$. 
Substituting these relations into Eqs.~(\ref{eqdlu})-(\ref{eqdphiu}), 
three perturbations $\vec{\cal X}={}^t(\delta \rho, \psi, \delta \phi)$ 
obey equation of the form (\ref{Xpereq}). 

Let us consider the radial propagation of dynamical perturbations 
for high frequency modes. The dispersion relation is given by 
Eq.~(\ref{dispr}), with the same values of $K_{21}$, $K_{31}$, 
$G_{21}$, $G_{31}$ and $K_{11}$, $G_{11}$ as those in 
Eqs.~(\ref{K210}), (\ref{K21}) and (\ref{K11}). 
Then, $\delta \rho$ propagates with the radial speed squared 
$c_{r1}^2=c_m^2$. An important difference from the $l \ge 2$ case 
is that the matrix components $G_{22}$ and $G_{32}$ 
vanish. Then, one of the solutions to Eq.~(\ref{dispr}) is 
$c_{r2}^2=0$, so that $\psi$ does not propagate.
The other solution associated with the propagation speed squared 
of $\delta \phi$ is 
\be
c_{r3}^2=\frac{K_{32}G_{23}-K_{22} G_{33}}
{fh (K_{22}K_{33}-K_{23}K_{32})}
={\frac { 4\phi'[ 8r^2 f h a_4 c_4 (\phi' a_1+ra_4)
-a_1^{2}{f}^{3/2}\phi' {\cal G} \sqrt {h}+2  ( a_1 f'+2 c_2 f ) a_4^{2}r^2] }
{f^{5/2}\sqrt {h}[(2{\cal P}_1-{\cal F})
h \mu^2 -2{\cal H} ^2 r^4 (\rho+P) ]}}\,,
\label{cr3l=0}
\ee
which is identical to Eq.~(\ref{cr}). 
Term $K_{22}K_{33}-K_{23}K_{32}$, which is associated with 
the no-ghost condition, is also the same as Eq.~(\ref{calK}). 
In summary, for $l=0$, the propagations of two dynamical perturbations 
$\delta \rho$ and $\delta \phi$ along the radial direction occur 
in the same manner as that for $l \geq 2$, with no propagation 
of the gravitational degree of freedom.

\subsection{$l=1$}

Let us proceed to the dipole mode ($l=1$). 
On using two relations $g_5=-2g_3$ and $g_6=-2g_4$ with $L=2$  
in Eq.~(\ref{eaction}), terms containing the dependence of 
$G$ and $K$ always appear as the derivatives of $G-K$.
We choose the gauges $h_0=0$, $K=0$, and $G=0$, 
but this does not completely fix the gauge degrees of freedom. 
Then, we impose the gauge condition $\delta \phi=0$ 
further. For $L=2$, the left hand-side of Eq.~(\ref{eqG}) vanishes identically, so 
we cannot exploit Eq.~(\ref{eqGu}) to eliminate $H_0$. 
Instead, we resort to Eq.~(\ref{eqH1}) together with 
Eq.~(\ref{delueq1}). Taking the time derivative of Eq.~(\ref{eqH1}) 
and using Eqs.~(\ref{eqB1})-(\ref{delueq1}), it follows that 
\be
\frac{1}{f} \left[ r^2 (\rho+P) \sqrt{fh}-2a_4 \right]\dot{H}_1
+\frac{2a_3}{f} \ddot{\psi}
-\frac{2a_4}{f} \ddot{h}_1+\frac{\sqrt{h}}{2f} 
r^2 (\rho+P) \left( f' \dot{v}-2f \dot{v}' \right)=0\,.
\label{eqH1f}
\ee
{}From Eq.~(\ref{dotveq}), we can express $\dot{v}$ and $\dot{v}'$ 
with respect to $\delta \rho$, $H_0$, $\delta \rho'$, and $H_0'$.

We solve Eqs.~(\ref{eqH0u}), (\ref{eqH2u}), (\ref{eqh1u}), and 
(\ref{eqH1f}) for $h_1$, $\dot{H}_1$, $H_0$, and 
$\ddot{h}_1$, respectively.
After taking the $r$ derivative of Eqs.~(\ref{eqH0u}) and (\ref{eqH2u}), 
we also express $h_1'$ and $H_0''$ with respect to 
other perturbations. 
Substituting all these relations into Eqs.~(\ref{eqdlu}) and (\ref{eqpsiu}), 
two dynamical perturbations $\vec{\cal X}={}^t(\delta \rho, \psi)$ obey 
equation of the form (\ref{Xpereq}) with $2 \times 2$ matrices 
${\bm K}$, ${\bm G}$, ${\bm Q}$, ${\bm M}$. 
For high-frequency modes the dispersion relation along the 
radial direction is given by Eq.~(\ref{dispr}).
Since $K_{21}=0$ and $G_{21}=0$ with the same values of 
$K_{11}$ and $G_{11}$ as those in Eq.~(\ref{K11}),  
$\delta \rho$ propagates with the sound speed 
squared $c_{r1}^2=c_m^2$. 
The propagation speed squared of $\psi$ is 
\be
c_{r3}^2=-\frac{G_{22}}{fh K_{22}}=
{\frac { 4\phi'[ 8r^2 f h a_4 c_4 (\phi' a_1+ra_4)
-a_1^{2}{f}^{3/2}\phi' {\cal G} \sqrt {h}+2  ( a_1 f'+2 c_2 f ) a_4^{2}r^2] }
{f^{5/2}\sqrt {h}[(2{\cal P}_1-{\cal F})
h \mu^2 -2{\cal H} ^2 r^4 (\rho+P) ]}}\,.
\label{cr3l=1}
\ee
This is equivalent to Eq.~(\ref{cr}), i.e., 
the sound speed squared of $\delta \phi$. 
The matrix component $K_{22}$ is also proportional to 
${\cal K}$ given by Eq.~(\ref{noghost}). 
Thus, for $l=1$, the matter and scalar-field perturbations propagate 
with the same forms of radial sound speeds as those derived for $l=0$. 
These results are consistent with those obtained in Ref.~\cite{Kobayashi:2014wsa}
without the perfect fluid.

%%%%%%%%%%%%%%%%%%%%%%%%%%%%%%%%%%%%%%%%%%%
\section{Concrete theories}
\label{theorysec}
%%%%%%%%%%%%%%%%%%%%%%%%%%%%%%%%%%%%%%%%%%%

Let us apply the stability conditions derived in Secs.~\ref{oddsec} 
and \ref{cssec} to two concrete theories in which hairy relativistic 
star solutions are known to exist. The first class is the case in which 
the nonminimal coupling $G_4(\phi)R$ is present.
The second class is a derivative coupling with the Einstein tensor
of the form $\phi\,G_{\mu \nu} \nabla^{\mu} \nabla^{\nu}\phi$.

\subsection{Nonminimally coupled theories}

The theories of spontaneous scalarization, $f(R)$ gravity, and 
Brans-Dicke theories can be accommodated by the action 
\be
{\cal S}=\int {\rm d}^4 x \sqrt{-g} \left[ 
G_4(\phi)R+G_2(\phi,X) \right]
+{\cal S}_m (g_{\mu \nu}, \Psi_m)\,.
\label{actp}
\ee
In the odd-parity sector, the stability conditions 
are ensured for
\be
{\cal H}={\cal F}={\cal G}=2G_4>0\,.
\label{HFG}
\ee
The radial and angular propagation speed squares for 
the odd-mode gravitational perturbation $\chi$ are 
\be
c_r^2=1\,,\qquad c_{\Omega}^2=1\,.
\ee
In the even-parity sector, the matter perturbation $\delta \rho$  
propagates with the radial and angular speed squares 
given by 
\be
c_{r1}^2=c_m^2\,,\qquad c_{\Omega 1}^2=c_m^2\,.
\ee
Using the background Eqs.~(\ref{back1}) and (\ref{back2}), 
the radial propagation speed squares (\ref{cr0}) and (\ref{cr}) of 
the perturbations $\psi$ and $\delta \phi$ reduce, respectively, to 
\ba
& &
c_{r2}^2=1\,,\\
& &
c_{r3}^2=\frac{\kappa+2XG_{2,XX}G_4}
{\kappa}\,,
\ea
where 
\be
\kappa \equiv G_{2,X}G_4+3G_{4,\phi}^2\,.
\ee
The no-ghost condition (\ref{noghost}) translates to 
${\cal K}=8 \phi'^2 h r^4 G_4 \kappa>0$. 
Under the condition (\ref{HFG}), this is satisfied for 
\be
\kappa>0\,.
\label{kappa}
\ee

Since $B_1=-1$ and $B_2=1$ for the theories under consideration, 
the angular propagation speed squares (\ref{cosq}) in the 
even-parity sector reduce to 
\be
c_{\Omega+}^2=c_{\Omega-}^2=1\,.
\ee
These results show that, apart from $c_{r3}^2$, all the other 
propagation speeds are equivalent to that of light. 
In the following, we consider more specific theories
that belong to the action (\ref{actp}).

\subsubsection{Theories of spontaneous scalarization}

In the Jordan frame, theories of spontaneous scalarization 
are given by the functions \cite{Kase:2020yhw,Kase:2020qvz}
\be
G_4(\phi)=\frac{M_{\rm pl}^2}{2}F(\phi)\,,\qquad
G_2(\phi,X)=\left( 1-\frac{3M_{\rm pl}^2 F_{,\phi}^2}{2F^2} 
\right) F(\phi)X\,,
\ee
where $M_{\rm pl}$ is the reduced Planck mass. 
For the realization of spontaneous scalarization, the nonminimal 
coupling $F(\phi)$ contains even power-law functions of $\phi$.
Since $G_{2,XX}=0$, it follows that $c_{r3}^2=1$. 
The no-ghost condition (\ref{kappa}) yields
\be
\kappa=\frac{M_{\rm pl}^2}{2}F^2 (\phi)>0\,,
\ee
which is automatically satisfied. 
Hence the odd- and even-parity stabilities of scalarized solutions
in theories of spontaneous scalarization are ensured 
under the condition (\ref{HFG}), i.e., 
\be
F(\phi)>0\,,
\ee
besides the condition $\rho+P>0$ in the matter sector. 
This main result is the same as that found in Ref.~\cite{Kase:2020qvz},
in spite of different values $c_{r1}^2$, $c_{r2}^2$, $c_{r3}^2$, and 
$c_{\Omega +}^2$.

\subsubsection{Brans-Dicke theories}

Brans-Dicke theories with a potential $V(\phi)$ are given 
by the functions \cite{Brans:1961sx}
\be
G_4(\phi)=\frac{1}{2}\phi\,,\qquad 
G_2(\phi,X)=\frac{\omega_{\rm BD}}{\phi}X-V(\phi)\,,
\ee
where $\omega_{\rm BD}$ is a constant, and we used 
the unit $M_{\rm pl}=1$.
Note that $f(R)$ gravity corresponds to the special 
case $\omega_{\rm BD}=0$. 
Since $G_{2,XX}=0$, we have $c_{r3}^2=1$. 
The stability conditions (\ref{HFG}) and (\ref{kappa}) reduce, respectively, to 
\be
2G_4=\phi>0\,,\qquad 
\kappa=\frac{1}{4} \left( 2\omega_{\rm BD}+3 \right)>0\,.
\ee
Under these conditions, 
there are neither ghost nor Laplacian instabilities for 
hairy relativistic star solutions in Brans-Dicke theories.

\subsection{Derivative coupling to Einstein tensor}

Theories of a derivative coupling to the Einstein tensor are given 
by the action \cite{Cisterna:2015yla,Cisterna:2015uya}
\be
{\cal S}=\int {\rm d}^4 x \sqrt{-g}\,\left[ \frac{1}{16\pi G_{\rm N}} R
-\frac{1}{2} \eta \phi G_{\mu \nu} \nabla^{\mu}\nabla^{\nu} \phi 
\right]+{\cal S}_m\,,
\label{NDCaction}
\ee
where $G_{\rm N}=(8 \pi M_{\rm pl}^2)^{-1}$ is Newton gravitational coupling, 
and $\eta$ is a constant.
The action (\ref{NDCaction}) corresponds to the functions 
$G_4=1/(16 \pi G_{\rm N})$ and $G_5(\phi)=-\eta \phi/2$. 
{}From Eq.~(\ref{back5}), the background scalar field obeys  
\be
\eta \sqrt{\frac{h}{f}} \phi' 
\left[ f(1-h)-rf' h \right]=0\,, 
\label{fhre}
\ee
where we imposed the boundary conditions 
$f \to 1$, $h \to 1$, and $\phi' \to 0$ 
at spatial infinity. 
Then, there is a nontrivial branch satisfying 
\be
f'=\frac{f(1-h)}{hr}\,.
\label{feqd}
\ee
Substituting Eq.~(\ref{feqd}) into Eq.~(\ref{back2}), we obtain
\be
\eta h \phi'^2=Pr^2\,.
\label{etare}
\ee
This means that the nontrivial branch (\ref{feqd}) corresponds 
to a solution where $\phi'$ does not vanish only inside 
the star ($P>0$). Substituting Eq.~(\ref{etare}) and its $r$ derivative as well as Eq.~(\ref{back4}) 
into Eq.~(\ref{back1}), it follows that 
\be
h'=\frac{1-h-4\pi G_{\rm N}r^2[(1+h)\rho+6h P]}
{r(1+4\pi G_{\rm N}r^2 P)}\,.
\label{heqd}
\ee
Solving Eqs.~(\ref{back4}), (\ref{feqd}), (\ref{etare}), and (\ref{heqd}) 
for a given NS equation of state, 
we obtain a hairy solution where the metric components 
are modified from those in GR inside 
the star \cite{Cisterna:2015yla,Cisterna:2015uya}. 
The stabilities of such hairy NS solutions were studied in Ref.~\cite{Kase:2020yjf}, 
but the treatment of how to integrate out the Schutz-Sorkin action was 
not appropriate as in Ref.~\cite{Kase:2020qvz}. 
In the following, we will reconsider the stabilities of hairy NS solutions 
satisfying the relation (\ref{etare}).

The stabilities of odd-parity perturbations require that 
\be
{\cal H}={\cal G}=\frac{1}{8\pi G_{\rm N}}+\frac{1}{2}Pr^2>0\,,\qquad 
{\cal F}=\frac{1}{8\pi G_{\rm N}}-\frac{1}{2}Pr^2>0\,.
\label{HGFcon}
\ee
For $P>0$ the former is trivially satisfied, while the latter gives 
$Pr^2<1/(4\pi G_{\rm N})$. 

Under the conditions (\ref{HGFcon}), the radial propagation speed squared 
$c_{r2}^2={\cal G}/{\cal F}$ in the even-parity sector is positive.
The other value (\ref{cr}) reduces to 
\be
c_{r3}^2=\frac{r^6 P^2 (1+4\pi G_{\rm N}r^2 P)}
{2\pi G_{\rm N}{\cal K}}\,,
\ee
where
\be
{\cal K}=\frac{r^6 P}{8\pi G_{\rm N}} \left[ (3h-1) \rho
(1+4\pi G_{\rm N}r^2 P)+4(27h-1)\pi G_{\rm N}r^2 P^2
+(19h-1)P \right]\,.
\label{calKde}
\ee
The absence of ghosts requires that ${\cal K}>0$, 
under which $c_{r3}^2 \geq 0$. 
Expanding Eq.~(\ref{calKde}) around $P=0$, it follows that 
\be
{\cal K}=\frac{r^6 (3h-1)\rho}{8 \pi G_{\rm N}}P+
{\cal O}(P^2)\,.
\label{calKde2}
\ee
Around the surface of NSs, the pressure $P$ approaches 0 
and hence the dominant contribution to ${\cal K}$ is the first 
term on the right hand-side of Eq.~(\ref{calKde2}). 
Then, we require the condition 
\be
h>\frac{1}{3}\,.
\label{hcon}
\ee
The explicit forms of $c_{\Omega \pm}^2$ are complicated, so 
we do not write their expressions here.
However, the expansion of Eq.~(\ref{cosq}) around $P=0$ leads to 
\ba
& &
c_{\Omega +}^2=1+{\cal O} (P^2)\,,\\
& &
c_{\Omega -}^2=-\frac{3(1-h^2)}{4h (3h-1)}+{\cal O} (P)\,.
\label{cOf}
\ea
The metric component $h$ is related to a positive mass function $M(r)$
as $h(r)=1-2G M(r)/r$, and hence $h<1$.
Under the no-ghost condition $h>1/3$ the leading-order contribution 
to Eq.~(\ref{cOf}) is negative, so that $c_{\Omega -}^2<0$.
Then, the angular Laplacian instability around the 
surface of star is inevitable for the hairy NS solution 
discussed above. 
This fact was first recognized in Ref.~\cite{Kase:2020yjf}, 
but it was based on the values of $c_{r1}^2$, $c_{r2}^2$, $c_{r3}^2$, ${\cal K}$, 
$c_{\Omega+}^2$, and $c_{\Omega-}^2$ derived by inappropriate 
treatment for the integration of the Schutz-Sorkin action. 
In spite of this difference, the leading-order term in Eq.~(\ref{cOf}) 
is exactly the same as that obtained in Ref.~\cite{Kase:2020yjf}, 
together with the no-ghost condition $h>1/3$.
Hence the angular Laplacian instability in the even-parity sector 
does not allow for the existence of hairy NS solutions 
in derivative coupling theories given by the action (\ref{NDCaction}). 

\section{Conclusions}

In this paper, we studied relativistic star perturbations on a static and 
spherically symmetric background in Horndeski theories 
given by the action (\ref{action}). 
We implemented a perfect fluid inside the star as a form 
of the Schutz-Sorkin action (\ref{SM}). 
We performed the analysis for the full Horndeski Lagrangian (\ref{LH}), 
while the previous studies \cite{Kase:2020qvz,Kase:2020yjf} 
restricted to a subclass of Horndeski theories. Moreover,
we decomposed the perturbations into odd- and even-parity modes 
arising from gravity, scalar-field, and matter sectors, and derived the full 
second-order actions without fixing particular gauges. 
In our formulation, we can choose any convenient gauges depending on the problems at hand. 
To our knowledge, this gauge-ready formulation was not addressed before 
even for a sub-class of Horndeski theories.

We first showed that the second-order action of odd-parity perturbations is of the form (\ref{Sodd}).
The reduced Lagrangian (\ref{Lodd}) contains three metric perturbations 
$W$, $Q$, $U$ and perturbations $\delta j$, $\delta {\cal A}$, $\delta {\cal B}$
arising from the perfect fluid. 
Choosing the gauge $U=0$ and using the perturbation equations in the matter 
sector, we showed that the dynamical perturbation $\chi=\dot{W}-Q'+2Q/r$ obeys 
Eq.~(\ref{chieq}). The field $\chi$ is coupled to a nondynamical fluid-velocity 
perturbation $\delta j$ satisfying $\dot{\delta j}=0$, whose property is 
consistent with the corresponding equation in GR \cite{Cunningham:1978zfa}. 
For the modes of high frequencies and large multipoles, 
the stability conditions in the odd-parity sector
are not affected by the presence of matter.

The second-order action of even-parity perturbations 
is given by Eq.~(\ref{Seven}), where the associated Lagrangian (\ref{eaction}) 
contains sixteen perturbed variables. 
We derived the full linear perturbation equations of motion without fixing gauges.
In Sec.~\ref{cssec}, we chose the gauge conditions $h_0=0$, $G=0$, and 
$K=0$ for the multipoles $l \geq 2$ and studied the propagation of even-parity modes.
Introducing a dynamical perturbation $\psi$ arising from the gravity sector 
and eliminating all nondynamical variables, the dynamical 
perturbations $\vec{{\cal X}}={}^t (\delta \rho, \psi, \delta \phi)$ 
satisfy closed-form equations of the form (\ref{Xpereq}). 
Among them, the matter perturbation is decoupled from others 
for high-frequency modes, so that the radial propagation speed 
squared of $\delta \rho$ is given by $c_{r1}^2=c_m^2$.
We also obtained the propagation speed squares of $\psi$ and $\delta \phi$ 
in the forms (\ref{cr0}) and (\ref{cr}), respectively.
While $c_{r2}^2$ coincides with the one derived in Ref.~\cite{Kobayashi:2014wsa}, 
$c_{r3}^2$ is subject to modification by the presence of matter. 
We also showed that, for both $l=0$ and $l=1$, the radial propagations of 
the perfect fluid and scalar field occur in the same manner as those for $l \geq 2$, with 
no propagation of the gravitational perturbation.

For large multipoles $l~(\gg 1)$, we derived the angular propagation speeds 
of even-parity perturbations by solving the dispersion relation (\ref{dispo}).
Like the radial mode, $\delta \rho$ is decoupled from $\psi$ and $\delta \phi$, 
so that the matter perturbation propagates with the speed squared 
$c_{\Omega 1}^2=c_m^2$ in the angular direction.
We also obtained the angular speed squares of $\psi$ and $\delta \phi$
in the forms (\ref{cosq}), where $B_1$ and $B_2$ are given by 
Eqs.~(\ref{B1def}) and (\ref{B2def}), respectively.
These values of $c_{\Omega \pm}^2$ were not derived in the literature 
even in the absence of matter. 
Given that there are some modified gravity (vector-tensor) theories in which 
hairy BH solutions are subject to Laplacian instabilities along the 
angular direction \cite{Kase:2018voo,Tsujikawa:2021typ}, 
our general expressions of $c_{\Omega \pm}^2$ will 
be useful to select stable NS and BH solutions in Horndeski theories. 

In Sec.~\ref{theorysec}, we applied our stability conditions to hairy 
NS solutions in several sub-classes of Horndeski theories. 
Note that similar investigations were performed in 
Refs.~\cite{Kase:2020qvz,Kase:2020yjf}, but the prescription 
of how to integrate out the Schutz-Sorkin action was not appropriate 
in these works. In Appendix B, we clarified this issue by 
deriving an effective matter action of the form (\ref{Lmeff}).
The analyses of Refs.~\cite{Kase:2020qvz,Kase:2020yjf} led to some 
different propagation speeds of even-parity perturbations 
in comparison to those in the current  paper.
On using the correct radial and angular propagation speeds, 
the main stability conditions discussed in Refs.~\cite{Kase:2020qvz,Kase:2020yjf} 
are hardly subject to modifications. 
For example, scalarized solutions in theories of spontaneous scalarization 
are stable under the condition $G_4(\phi)>0$.  
Hairy solutions in theories of the derivative coupling to the Einstein tensor
are subject to the angular Laplacian instability around 
the surface of NSs. 
In particular, the latter fact shows the importance of scrutinizing 
the stabilities of NSs and BHs against both odd- and even-parity 
perturbations before computing quasi-normal modes 
of odd-parity perturbations alone \cite{Blazquez-Salcedo:2018tyn,AltahaMotahar:2019ekm}. 

It will be of interest to judge the stabilities of hairy NS and BH 
solutions by using our general conditions derived in this paper. 
The applications to the tidal deformation of a NS-NS binary system 
and to the calculations of NS and BH quasi-normal frequencies 
will be the next important steps to probe the physics of strong 
gravity regimes through the observations of GWs.

%%%%%%%%%%%%%%%%%%
\section*{Acknowledgements}
%%%%%%%%%%%%%%%%%%

We thank Rampei Kimura, Kei-ichi Maeda, and Seiga Sato for useful discussions.
RK is supported by the Grant-in-Aid for Young Scientists B 
of the JSPS No.~20K14471. 
ST is supported by the Grant-in-Aid for Scientific Research 
Fund of the JSPS No.~19K03854.

\renewcommand{\theequation}{A.\arabic{equation}}
\setcounter{equation}{0}

%%%%%%%%%%%%%%%
\section*{Appendix A: Coefficients in the second-order action of 
even-parity perturbations}
\label{app:coefficients}
%%%%%%%%%%%%%%%

The coefficients of the reduced Lagrangian \eqref{eaction} of 
even-parity perturbations are 
\ba
a_1&=&\sqrt{fh} \left[  \left\{ G_{4,\phi}+\frac12 h ( G_{3,X}-2 G_{4,\phi X} ) \phi'^2 \right\} r^2
+2 h \phi' \left\{ G_{4,X}-G_{5,\phi}-\frac12h ( 2 G_{4,XX}-G_{5,\phi X} ) \phi'^2 \right\} r
\right.
\notag\\
&&
\left.
+\frac12 G_{5,XX} h^3\phi'^4-\frac12 G_{5,X} h ( 3 h-1 ) \phi'^2 \right]\,, \notag\\
a_2&=&\sqrt{fh}\left( {\frac {a_1}{\sqrt{fh}}} \right)' 
- \left( {\frac {\phi''}{\phi'}}-\frac12 {\frac {f'}{f}} \right) a_1
+{\frac {r}{\phi'} \left( {\frac {f'}{f}}-{\frac {h'}{h}} \right) a_4}
-\frac12 {\frac { \sqrt {f}r^2( \rho+P )}{\sqrt {h}\phi'}}\,,\notag\\
a_3&=&-\frac12 \phi' a_1-ra_4\,,\qquad
a_4=\frac{\sqrt{fh}}{2} {\cal H}\,, \qquad
a_5=a_2'-a_1''\,,\qquad
a_6=- {\frac {\sqrt {f}}{2\sqrt {h}\phi'} 
\left( {\cal H}' + \frac{{\cal H}}{r}-{\frac {{\cal F}}{r}} \right) }\,,
\notag\\
a_7 &=&a_3'+\frac14 {\frac {\sqrt {f}r^2( \rho+P )}{\sqrt {h}}}\,,\qquad
a_8=-\frac12 {\frac {a_4}{h}}\,,\qquad
a_9= a_4'+ \left( \frac1r-\frac12 {\frac {f'}{f}} \right) a_4\,,\qquad
a_{10}=\frac12 {\frac {r^2\sqrt {f}}{\sqrt {h}}}\,,\notag\\
b_1&=& {\frac {1}{2f}}a_4\,,\qquad
\tilde{b}_1=-{\frac {h r^2}{f}}f_1\,,\qquad
b_2=-{\frac {2}{f}}a_1\,,\qquad
b_3=- {\frac {2}{f}}(a_2-a_1')\,,\qquad
b_4=-{\frac {2}{f}}a_3\,,\qquad
b_5=-2 b_1\,,\notag\\
c_1&=&-{\frac {1}{fh}}a_1\,,\notag\\
c_2&=&\sqrt{fh} \left[  \left\{  
\frac{1}{2f}\left( -\frac12 h ( 3 G_{3,X}-8 G_{4,\phi X} ) \phi'^2
+\frac12 h^2 ( G_{3,XX}-2 G_{4,\phi XX} ) \phi'^4
-G_{4,\phi} \right) r^2
\right.\right.
\notag\\
&&
\left.\left.
-{\frac {h\phi'}{f}} \left( 
\frac12 {h^2 ( 2 G_{4,XXX}-G_{5,\phi XX} ) \phi'^4}
-\frac12 {h ( 12 G_{4,XX}-7 G_{5,\phi X} ) \phi'^2}
+3 ( G_{4,X}-G_{5,\phi} ) \right) r
\right.\right.
\notag\\
&&
\left.\left.
+\frac{h\phi'^2}{4f}\left(
G_{5,XXX} h^3\phi'^4
- G_{5,XX} h ( 10 h-1 ) \phi'^2
+3 G_{5,X}  ( 5 h-1 ) 
\right) \right\} f'
\right.
\notag\\
&&
\left.
+\phi' \left\{ \frac12G_{2,X}-G_{3,\phi}
-\frac12 h ( G_{2,XX}-G_{3,\phi X} ) \phi'^2 \right\} r^2
\right.
\notag\\
&&
\left.
+ 2\left\{ -\frac12h ( 3 G_{3,X}-8 G_{4,\phi X} ) \phi'^2
+\frac12h^2 ( G_{3,XX}-2 G_{4,\phi XX} ) \phi'^4
-G_{4,\phi} \right\} r
\right.
\notag\\
&&
\left.
-\frac12 h^3 ( 2 G_{4,XXX}-G_{5,\phi XX} ) \phi'^5
+\frac12 h \left\{ 2\left(6 h-1\right) G_{4,XX}+\left(1-7 h\right)G_{5,\phi X} \right\} \phi'^3
- ( 3 h-1 )  ( G_{4,X}-G_{5,\phi} ) \phi' \right] \,,\notag\\
c_3&=&-\frac12 {\frac {\sqrt {f} r^2}{\sqrt {h}}}\frac{\partial{\cal E}_{11}}{\partial\phi}\,,\notag\\
c_4&=&\frac14 \frac {\sqrt {f}}{\sqrt {h}} 
\left[ {\frac {h\phi'}{f} \left\{ 
2 G_{4,X}-2 G_{5,\phi}
-h ( 2 G_{4,XX}-G_{5,\phi X} ) \phi'^2
-{\frac {h\phi'  ( 3 G_{5,X}-G_{5,XX} \phi'^2h ) }{r}} \right\}}f'
\right.
\notag\\
&&
\left.
+4 G_{4,\phi}
+2 h ( G_{3,X}-2 G_{4,\phi X} ) \phi'^2
+{\frac {4 h ( G_{4,X}-G_{5,\phi} ) \phi'-2 h^2 ( 2 G_{4,XX}-G_{5,\phi X} ) \phi'^3}{r}} \right] \,,\notag\\
c_5&=&-h \phi'c_4-\frac12 {\frac {\sqrt{fh}}{r}}{\cal G}-\frac12 {\frac {f'}{f}}a_4\,,\notag\\
c_6&=&\frac18 {\frac {f' \phi' }{f}}a_1+\frac12 {\frac {f' r}{f}}a_4-\frac14 \phi' c_2+\frac12 h\phi' rc_4
+\frac14 \sqrt{fh}\,{\cal G}
\,,\notag\\
d_1&=&{\frac {1}{2f}}a_4\,,\qquad
d_2=2 hc_4\,,\notag\\
d_3&=&
-{\frac {1}{r^2} \left( {\frac {2\phi''}{\phi'}}+{\frac {h'}{h}} \right) }a_1
+{\frac {2f}{ ( f' r-2 f ) \phi'} \left( 
{\frac {2\phi''}{h\phi' r}}
+ {\frac {{f'}^{2}}{f^2}}
- {\frac {f' h'}{fh}}
-{\frac {2f'}{fr}}
+{\frac {2h'}{hr}}
+ {\frac {h'}{h^2r}} \right) }a_4
\notag\\
&&
+{\frac {f' r-2 f}{fr}}\frac{\partial a_4}{\partial\phi}
+{\frac {\sqrt {f}}{\phi' \sqrt {h}r^2}}{\cal F}
-{\frac {{f}^{3/2}}{\sqrt {h} ( f' r-2 f ) \phi'} 
\left( {\frac {f'}{fr}}+{\frac {2\phi''}{\phi' r}}+{\frac {h'}{hr}}-\frac{2}{r^2} \right) }{\cal G}
-{\frac {\sqrt {f} ( \rho+P ) }{\phi' \sqrt {h}}}\,,\notag\\
d_4&=&\frac12 {\frac {\sqrt{fh}}{r^2}}{\cal G}
\,,\notag\\
e_1&=&{\frac {1}{\phi' fh} \left[  \left( {\frac {f'}{f}}+\frac12 {\frac {h'}{h}} \right) a_1
-2 a_1'+a_2-2 rha_6 \right] }\,,\notag\\
e_2&=&-\frac{1}{2\phi'} \left(\frac{f'}{f}a_1+2 c_2+4 hrc_4\right)\,,\qquad
e_3=\frac14 {\frac {\sqrt {f}r^2}{\sqrt {h}}}\frac{\partial{\cal E}_{\phi}}{\partial\phi}\,,\notag\\
e_4&=&{\frac {1}{\phi'}}c_4'-\frac12 {\frac {f' }{f\phi'^2h}}a_4'
-\frac12 {\frac {\sqrt {f}}{\phi'^2\sqrt {h}r}}{\cal G}'
+{\frac {1}{h\phi' r^2} \left( {\frac {\phi''}{\phi'}}+\frac12 {\frac {h'}{h}} \right) }a_1
\notag\\
&&
+{\frac {1}{4h\phi'^2} \left[ {\frac { ( f' r-6 f ) f'}{f^2r}}
+\frac {h'  ( f' r+4 f ) }{hrf}
-{\frac {4f ( 2 \phi'' h+h' \phi' ) }{\phi' h^2r ( f' r-2 f ) }} \right] }a_4
+\frac12 {\frac {h'}{h\phi'}}c_4
-\frac12 {\frac {f' r-2 f}{fhr\phi'}}\frac{\partial a_4}{\partial \phi}
\notag\\
&&
+\frac12 {\frac {f' hr-f}{r^2\sqrt {f}\phi'^2{h}^{3/2}}}{\cal F}
+\frac12 {\frac {\sqrt {f}}{r\phi'^2{h}^{3/2}} 
\left[ {\frac {f ( 2 \phi'' h+h' \phi' ) }{h\phi'  ( f' r-2 f ) }}+\frac12 {\frac {2 f-f' hr}{fr}} \right] }{\cal G}
+\frac12 {\frac {\sqrt {f} ( \rho+P ) }{{h}^{3/2}\phi'^2}}
\,,\notag\\
f_1&=&\frac12 {\frac {\sqrt {f} ( \rho+P ) }{\sqrt {h}}}\,,\qquad
f_2=-\frac12 {\frac {r^2\sqrt {f}\,c_m^2}{\sqrt {h} ( \rho+P ) }}\,,\qquad
f_3=\sqrt {h}nr^2\,,\qquad
f_4={\frac {n r^2}{\sqrt {h} ( \rho+P ) }}\,,\qquad
f_5=\frac12 {\frac {f_3}{h}}\,,\notag\\
f_6&=& \sqrt {f}f_3\,,\qquad
f_7=\frac12 {\frac {\sqrt {h}nr^2 ( c_m^2-1 ) f'}
{\sqrt {f}c_m^2}}+\frac12 {\frac {r^2\sqrt {f}h'n}{\sqrt {h}}}
+2 r\sqrt{fh}\,n\,,
\qquad
f_8=-{\frac {\sqrt {f}}{hr^2}}f_3\,,\notag\\
f_9&=&h r^2f_1\,,\qquad
f_{10}=-\sqrt {f}f_3\,,\qquad
f_{11}=-{\frac {\sqrt {f}}{hr^2}}f_3\,,\qquad
f_{12}=f_3\,,\qquad
f_{13}=-{\frac {1}{h}}f_3\,,\notag\\
g_1&=&{\frac {r^2}{f}}a_4\,,\qquad
g_2=-\frac12 r^2a_4\,,\qquad
g_3=-\frac18 {\frac {r^2}{\sqrt{fh}}}{\cal F}\,,\qquad
g_4=\frac18 r^2\sqrt{fh}{\cal G}\,,\qquad
g_5=-2g_3\,,\notag\\
g_6&=&-2g_4\,,\qquad
g_7=\frac12 {\frac {r^2}{fh}}a_4\,,\qquad
g_8=-{\frac {r^2}{f}}a_6\,,\qquad
g_9=-\frac12 {\frac {r ( f' r-2 f ) }{f^2}}a_4\,,\qquad
g_{10}=- {\frac {1}{2h}}f_3\,,\notag\\
g_{11}&=&{\frac {1}{2\sqrt{fh}}}{\cal F}\,,\qquad
g_{12}=-hr^2c_4\,,\qquad
g_{13}=\frac14 {\frac {r^2f'}{f}}a_4-\frac12 r^2a_4'-\frac32 ra_4\,,\qquad
g_{14}=-\frac12 r^2c_5\,,\notag\\
g_{15}&=&-\frac12 r^2d_3\,,\qquad
g_{16}=-\frac12 \sqrt{fh}\,{\cal G}\,,\notag\\
k_1&=&-\frac14 {\frac { \sqrt {f}}{\sqrt {h}}}{\cal F}\,,\qquad
k_2=\frac14 {\frac {\sqrt {f}}{\sqrt {h}}}{\cal G}\,,\qquad
k_3=\frac{2f{\cal G}'+f'({\cal G}-{\cal F})}{4\sqrt{fh}\,\phi'}
\,,\notag\\
m_1&=& {\frac {1}{2f}}a_4\,,\qquad
m_2=-{\frac {1}{f}}a_4\,,\qquad
m_3={\frac {1}{fr}}a_4'
+{\frac {1}{fr^2}}\left(1-\frac{rf'}{f}\right)a_4
-{\frac {1}{2\sqrt {fh}\,r^2}}{\cal F}\,,\qquad
m_4=-{\frac {1}{fh}}a_4\,,\notag\\
m_5&=&{\frac {2}{fr}}a_4\,,\qquad
m_6={\frac {2}{f}}a_6\,,\qquad
m_7={\frac {f' }{f^2}}a_4\,,\qquad
m_8={\frac {1}{hr^2}}f_3\,,
\ea
where ${\cal H}$, ${\cal F}$, ${\cal G}$, ${\cal E}_{11}$, and 
${\cal E}_{\phi}$ are given, respectively, by 
Eqs.~(\ref{cHdef}), (\ref{cFdef}), (\ref{cGdef}), (\ref{back2}), and (\ref{back5}). 

%%%%%%%%%%%%%%%%%%%%%%%%%%%%%%
\section*{Appendix B: Effective matter Lagrangian in the even-parity sector}
\label{matterac}
%%%%%%%%%%%%%%%%%%%%%%%%%%%%%%

We derive the effective matter Lagrangian for even-parity perturbations by 
integrating out most of the nondynamical variables.
{}From Eq.~(\ref{eaction}), the Lagrangian containing the matter 
perturbations $v$, $\delta {\cal A}_1$, $\delta {\cal A}_2$, 
$\delta {\cal B}_1$, $\delta {\cal B}_2$, $\delta u_r$, $\delta \ell$, 
and $\delta \rho$ (with explicit coefficients) is given by 
\ba
{\cal L}_m &=& 
\frac{L \sqrt{f} (\rho+P)}{2\sqrt{h}} v^2
+\frac{n r^2}{\sqrt{h}} \left( h H_1-\dot{\delta {\cal B}}_1 \right) 
\delta {\cal A}_1+\frac{L n(h_0-\sqrt{f} v-r^2\dot{\delta {\cal B}}_2)}{\sqrt{h}}
\delta {\cal A}_2+\frac{r^2\sqrt{fh} (\rho+P)}{2} \delta u_r^2 \nonumber \\
&&-nr^2 \sqrt{fh} (\delta {\cal A}_1+\delta\ell')\delta u_r
+\frac{nr^2}{\sqrt{h}} \left[ \frac{\dot{\delta \rho}}{\rho+P}
+\dot{K}+\frac{\dot{H}_2}{2}-\frac{L \dot{G}}{2}+
\frac{L}{r^2} (h_0-\sqrt{f}v) \right] \delta \ell
+n r^2 \sqrt{h}\,H_1 \delta \ell' \nonumber \\
&& -\frac{r^2 \sqrt{f}\,c_m^2}{2\sqrt{h}(\rho+P)} \delta \rho^2
+\frac{r^2 \sqrt{f}}{2\sqrt{h}}H_0 \delta \rho\,,
\label{Lmf}
\ea
where we integrated the term 
$n r^2 \sqrt{fh}\,\delta \ell \delta u_r'$ by parts.
As we mentioned at the end of Sec.~\ref{peroddsec}, 
varying the Lagrangian (\ref{Lmf}) with respect to the perturbations
with only its linear dependence and substituting the resulting equation into Eq.~(\ref{Lmf}) 
does not generally give rise to a correct reduced Lagrangian. 
In Ref.~\cite{Kase:2020qvz}, the authors varied the Lagrangian (\ref{Lmf}) 
with respect to $\delta {\cal B}_1$ and $\delta {\cal B}_2$, and 
plugged the resulting equations $\delta\dot{\cal A}_1=0$ and 
$\delta\dot{\cal A}_2=0$ into Eq.~(\ref{Lmf}) in order to eliminate the 
terms proportional to $\dot{\delta {\cal B}}_1 \delta {\cal A}_1$ 
and $\dot{\delta {\cal B}}_2 \delta {\cal A}_2$. 
However, this process finally leads to an inconsistent value 
$c_{r1}^2=0$ for the radial matter sound speed squared.

To obtain the reduced matter Lagrangian correctly, we first need to vary 
Eq.~(\ref{Lmf}) with respect to perturbations containing the squared dependence 
in addition to the linear dependence. The quantity $v$ is such a variable, 
so the variation of Eq.~(\ref{Lmf}) with respect to $v$ gives 
\be
v=\frac{n}{\rho+P} \left( \delta \ell+\delta {\cal A}_2 \right)\,,
\label{vApe1}
\ee
which corresponds to Eq.~(\ref{soldl}). Now, we can substitute 
Eq.~(\ref{vApe1}) into Eq.~(\ref{Lmf}) to eliminate $v$.
After this procedure the Lagrangian contains the terms 
proportional to $\delta {\cal A}_2^2$ as well as $\delta {\cal A}_2$, 
so the variation with respect to $\delta {\cal A}_2$ leads to 
\be
\delta {\cal A}_2=-\delta \ell-\frac{\rho+P}{n\sqrt{f}}
\left( r^2 \dot{\delta {\cal B}}_2-h_0 \right)\,.
\label{cAApe1}
\ee
On using Eq.~(\ref{vApe1}), it is clear that the relation (\ref{cAApe1}) 
is consistent with (\ref{eqA2}). 
Substituting Eq.~(\ref{cAApe1}) into the Lagrangian (derived after the 
substitution of $v$) gives rise to
the term proportional to $\dot{\delta {\cal B}}_2^2$, so the variation with 
respect to $\delta {\cal B}_2$ yields
\be
\dot{\delta {\cal B}}_2=\frac{h_0}{r^2}
-\frac{n\sqrt{f}}{r^2 (\rho+P)} \delta \ell+{\cal C}(r)\,,
\label{cBApe1}
\ee
where ${\cal C}(r)$ is a function of $r$. 
Then, from Eqs.~(\ref{cAApe1}) and (\ref{cBApe1}), it follows that 
\be
\delta {\cal A}_2=-\frac{r^2 (\rho+P)}{n\sqrt{f}}{\cal C}(r)\,.
\ee
The time derivative of this relation is consistent with Eq.~(\ref{eqB2}). 
When we derived the second-order differential equation (\ref{eqdlu}) 
of $\delta \rho$, we took the time derivative of Eq.~(\ref{eqdl}) 
and eliminated the term $\dot{\delta u}_r$ by using Eq.~(\ref{eqdur}).
In this procedure, the $\delta {\cal A}_2$-dependent term 
vanishes due to the relation (\ref{eqB2}). 
Since the arbitrary function ${\cal C}(r)$ does not affect the 
dynamics of matter perturbations $\delta \rho$, we can set 
\be
{\cal C}(r)=0 \quad \to \quad  
\delta {\cal A}_2=0\,,
\label{A20}
\ee
without the loss of generality. 
Then, the term $\delta {\cal A}_2$ vanishes from the reduced Lagrangian.

The next step is to vary the reduced Lagrangian further 
with respect to $\delta u_r$. 
This leads to 
\be
\delta u_r=\frac{n}{\rho+P} \left( \delta \ell'
+\delta {\cal A}_1 \right)\,.
\label{durAp}
\ee
Taking the $r$ derivative of Eq.~(\ref{soldl}) and eliminating 
the term $\delta \ell'$ from Eq.~(\ref{durAp}), we find that Eq.~(\ref{durAp}) 
is equivalent to Eq.~(\ref{eqdur}). 
On using Eq.~(\ref{durAp}), the reduced Lagrangian possesses
the term proportional to $\delta {\cal A}_1^2$, so the variation 
with respect to $\delta {\cal A}_1$ gives 
\be
\delta {\cal A}_1=-\delta \ell' 
+\frac{\rho+P}{n\sqrt{f}\,h} \left( h H_1-\dot{\delta {\cal B}}_1 
\right)\,,
\label{A1Ap}
\ee
which is consistent with Eq.~(\ref{eqA2}) on account of Eq.~(\ref{durAp}).
After eliminating $\delta {\cal A}_1$ from the Lagrangian, there are terms 
proportional to $\dot{\delta {\cal B}}_1^2$, so the variation with respect to 
$\delta {\cal B}_1$ leads to 
\be
\dot{\delta {\cal B}}_1=h H_1-\frac{n \sqrt{f}\,h}{\rho+P}
\delta \ell'+{\cal D}(r)\,,
\label{B1Ap}
\ee
where ${\cal D}(r)$ is a function of $r$. 
Substituting Eq.~(\ref{B1Ap}) into Eq.~(\ref{A1Ap}), we have 
\be
\delta {\cal A}_1=-\frac{\rho+P}{n\sqrt{f}\,h}{\cal D}(r)\,,
\ee
whose time derivative is consistent with Eq.~(\ref{eqB1}). 
The perturbation $\delta {\cal A}_1$ vanishes from 
the second-order differential equation (\ref{eqdlu})  
of $\delta \rho$, 
so we can set 
\be
{\cal D}(r)=0 \quad \to \quad  
\delta {\cal A}_1=0\,.
\label{Drre}
\ee
On using Eq.~(\ref{B1Ap}) with Eq.~(\ref{Drre}), we end up 
with the reduced Lagrangian containing $\delta \rho$, 
$\delta \ell$, and their derivatives. 
{}From Eqs.~(\ref{vApe1}) and (\ref{A20}) we have 
$\delta \ell=(\rho+P)v/n$, so $\delta \ell$ and $\delta \ell'$ 
can be replaced with $v$ and $v'$. 
Then, the reduced effective Lagrangian for the perfect fluid
is given by 
\ba
{\cal L}_m &=& \frac{r^2}{\sqrt{h}} \left[ v \dot{\delta \rho}
+\frac{\sqrt{f}}{2} H_0 \delta \rho-\frac{\sqrt{f}}{2(\rho+P)}
c_m^2 \delta \rho^2 \right]
-\frac{\rho+P}{8 f^{3/2}\sqrt{h}} \left[4L f^2 v^2
+r^2 h (2fv'-f' v)^2 \right] \nonumber\\
& &+\frac{\rho+P}{2f \sqrt{h}} 
\left[ 2L f h_0 v-r^2 f (L\dot{G}-\dot{H}_2-2\dot{K})v
-r^2f' h H_1 v+2r^2 fh H_1 v' \right]\,.
\label{Lmeff}
\ea
Varying (\ref{Lmeff}) with respect to $v$ and $\delta \rho$, 
we obtain the perturbation equations same as 
Eqs.~(\ref{eqdl}) and (\ref{eqdrho}), respectively. 
For the purpose of deriving the perturbation equations, 
we can directly employ the reduced matter Lagrangian (\ref{Lmeff})
besides the second-order Lagrangians of gravitational and scalar-field
perturbations.

%%%%%%%%%%%%%%%%

\end{document}